\documentclass{article}

\usepackage{PRIMEarxiv}

\usepackage[utf8]{inputenc} %
\usepackage[T1]{fontenc}    %
\usepackage{hyperref}       %
\usepackage{url}            %
\usepackage{booktabs}       %
\usepackage{amsfonts}       %
\usepackage{nicefrac}       %
\usepackage{microtype}      %
\usepackage{lipsum}
\usepackage{graphicx}
\graphicspath{{media/}}     %
\usepackage{graphicx}%
\usepackage{multirow}%
\usepackage{multicol}
\usepackage{amsmath,amssymb,amsfonts}%
\usepackage{amsthm}%
\usepackage{mathrsfs}%
\usepackage[title]{appendix}%
\usepackage{xcolor}%
\usepackage{textcomp}%
\usepackage{manyfoot}%
\usepackage{booktabs}%
\usepackage{algorithm}%
\usepackage{algorithmicx}%
\usepackage{algpseudocode}%
\usepackage{listings}%

\usepackage{colortbl}
\graphicspath{{./Figures/}{../Figures/}}

\title{Delay Parameter Selection in Permutation Entropy Using Topological Data Analysis
\thanks{Myers, A.D., Chumley, M.M. \& Khasawneh, F.A. Delay Parameter Selection in Permutation Entropy Using Topological Data Analysis. La Matematica (2024). https://doi.org/10.1007/s44007-024-00110-4} 
}

\author{
  Audun D. Myers \\
  Pacific Northwest National Laboratory \\
  Richland WA\\
  \texttt{audun.myers@pnnl.gov} \\
   \And
  Max M. Chumley,~Firas A. Khasawneh\thanks{Address all correspondence to this author.
} \\
  Department of Mechanical Engineering\\
  Michigan State University \\
  East Lansing MI\\
  \texttt{chumleym@msu.edu, khasawn3@msu.edu} \\
}

\begin{document}
\maketitle

\begin{abstract}
Permutation Entropy (PE) is a powerful tool for quantifying the complexity of a signal which includes measuring the regularity of a time series. Additionally, outside of entropy and information theory, permutations have recently been leveraged as a graph representation, which opens the door for graph theory tools and analysis.
Despite the successful application of permutations in a variety of scientific domains, permutations requires a judicious choice of the delay parameter $\tau$ and dimension $n$. 
However, $n$ is typically selected within an accepted range giving optimal results for the majority of systems. Therefore, in this work we focus on choosing the delay parameter, while giving some general guidance on the appropriate selection of $n$ based on a statistical analysis of the permutation distribution.
Selecting $\tau$ is often accomplished using trial and error guided by the expertise of domain scientists. 
However, in this paper, we show how persistent homology, a commonly used tool from  Topological Data Analysis (TDA), provides methods for the automatic selection of $\tau$.
We evaluate the successful identification of a suitable $\tau$ from our TDA-based approach by comparing our results to both expert suggested parameters from published literature and optimized parameters (if possible) for a wide variety of dynamical systems.
\end{abstract}


\section{Introduction}\label{intro}
Shannon entropy, which was introduced in 1948 \cite{Shannon1951}, is a summary  statistic measuring the regularity of a dataset. 
Since then, several new forms of entropy have been popularized for time series analysis. 
Some examples include approximate entropy~\cite{Pincus1991}, sample entropy~\cite{Richman2000}, and permutation entropy (PE)~\cite{Bandt2002}. 
While all of these methods measure the regularity of a sequence, PE does this through the motifs or permutations found within the signal. This allows for PE to be related to predictability~\cite{Garland2014, Pennekamp2019}, which is useful for detecting dynamic state changes.
Similar to Shannon Entropy, PE~\cite{Bandt2002} is quantified as the summation of the probabilities of a data type (see Eq.~\eqref{eq:PE}), where the data types for PE are permutations (see Fig.~\ref{fig: Permutations_n6}), which we represent as $\pi$. Permutations have recently been used in other applications such as ordinal partition networks~\cite{McCullough2015, Myers2019} and the conditional entropy of these networks~\cite{Shahriari2020}. The permutation parameters $n$ and $\tau$ represent the permutation size and spacing, respectively. More specifically, $\tau$ is the embedding delay lag applied to the series and $n$ is a natural number that describes the dimension of the permutation. 
In this study we focus on selecting $\tau$ using methods based in Topological Data Analysis (TDA) since the dimension is typically chosen in the range $3<n\leq 7$ for most applications \cite{Riedl2013}. However, we still provide a novel and simple guidance on the automatic selection of $n$ based on a statistical analysis of the permutations.

Currently, the most common method for selecting PE parameters is to adopt the values suggested by domain scientists. 
For example, Li et al.~\cite{Li2013} suggest using $\tau=3$ and $n=3$ for electroencephalographic (EEG) data, Zhang and Liu~\cite{Zhang2018} suggest $\tau = 3$ or $5$ and $n\in[3,5]$ for logistic maps, and Frank et al.~\cite{Frank2006} suggest $\tau = 2$ or $3$ and $n\in[3,7]$ for heart rate applications. One main disadvantage of using suggested parameter values for an application is the high dependence of PE on the sampling frequency. 
As an example, Popov et al.~\cite{popov2013permutation} showed the importance of considering the sampling frequency when selecting $\tau$ for an EEG signal. 
Another limitation is the need for application expertise in order to determine the needed parameters. This can hinder using PE in new applications that have not been sufficiently explored. Consequently, there is a need for an automatic, application-independent parameter selection algorithm for PE.

Several methods have been developed for estimating $\tau$ for phase space reconstruction via Takens embedding~\cite{Takens1981}.
Some of which include mutual information~\cite{Fraser1986}, autocorrelation~\cite{grassberger1983measuring}, and phase space methods~\cite{buzug1992optimal}.
There has also been recent work in determining if these methods are suitable for the delay parameter selection for permutations~\cite{Myers2020}. Outside of this, a general framework for selecting both $n$ and $\tau$ was introduced in~\cite{Riedl2013}. In this manuscript our goal is to determine other, TDA-based methods for selecting $\tau$ for PE and to draw a connection between permutations and state space reconstruction.
Specifically, in Section~\ref{ssec:relating_PE_to_TDA}, we relate permutations to state space reconstruction to provide a justification for using the lag parameter from the latter to select $\tau$ for the former.
We then present a novel TDA-based tool for finding $\tau$.  
For our approach we compute the $0$-D sublevel set persistence in both time and frequency domains to obtain approximations of the maximum significant frequency.
We then utilize Nyquist's sampling theorem to find an appropriate $\tau$ value.

To determine the viability of our methods, PE parameters are generated and compared to expert suggested values, optimal parameters based on maximizing the difference between permutation entropy for periodic and chaotic signals, and the delay corresponding to the first minima of mutual information. The method of mutual information was chosen as a basis for comparison based on its accuracy in selecting $\tau$ for PE as demonstrated in~\cite{Myers2020}.

The paper is organized as follows. 
In Section \ref{sec:PE_example} we describe PE and show its computation using a simple example. 
Section \ref{sec:tda_overview} provides an overview of the tools that we use from TDA. In Section~\ref{sec:timeLag}, we review the method of mutual information for embedding delay parameter selection (Section \ref{sec:MI}) and the $0$-D sublevel set persistence methods for selecting $\tau$ (Section \ref{ssec:0d_sublevel_for_delay}). In Section~\ref{sec:dimension} we introduce our novel approach for selecting the permutation dimension $n$ using the statistics of permutations and highlight the comparison to standard Takens' embedding techniques which, from our analysis, are not appropriate for permutation dimension selection.
The results of each method for a variety of systems are then presented in Section \ref{sec:results}, and the concluding remarks are presented in Section~\ref{sec:conclusions}.

\subsection{Permutation Entropy Example}
\label{sec:PE_example}
\begin{figure}[h] 
    \centering
    \includegraphics[width=0.35\textwidth]{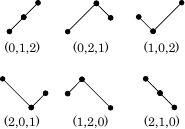}
    \caption{All possible permutation configurations (motifs) for n = 3, where $[\pi_1 \ldots \pi_6] = [(0, 1, 2) \ldots (2,1,0)]$.}
    \label{fig: Permutations_n6}
\end{figure}

Permutation entropy $H(n)$ for permutation dimension $n$ is calculated according to \cite{Bandt2002} as
\begin{equation}
    H(n) = -\sum{} p(\pi_i) \log{p(\pi_i)},
    \label{eq:PE}
\end{equation}
where $p(\pi_i)$ is the probability of a permutation $\pi_i$, and $H(n)$ has units of bits when the logarithm is of base $2$. 
The permutation entropy parameters $\tau$ and $n$ are used when selecting the permutation size: $\tau$ is the number of time steps between two consecutive points in a uniformly sub-sampled time series, and $n$ is the permutation length or motif dimension.
Using a real-valued data set $X$ and a measurement of the set $x_i \in X$, we can define the vector $v_i = [x_{i}, x_{i+\tau}, x_{i+2\tau},\ldots ,x_{i+(n-1)\tau}]$, which has the permutation $\pi_i$.
To better understand the possible permutations, consider an example with third degree ($n = 3$) permutations. 
This results in six possible motifs as shown in Fig.~\ref{fig: Permutations_n6}. 

Next, to further demonstrate PE with an example, consider the sequence $X = [4, 7, 9, 10, 6, 11, 3]$ with third order permutations $n=3$ and time delay $\tau = 1$. 
The sequence can be broken down into the following permutations: two $(0, 1, 2)$ permutations, one $(1,0,2)$ permutation, and two $(1, 2, 0)$ permutations for a total of 5 permutations. 
Applying Eq.~\eqref{eq:PE} yields
\begin{equation*}
   H(3) =  -\frac{2}{5}\log{\frac{2}{5}} -\frac{2}{5}\log{\frac{2}{5}} -\frac{1}{5}\log{\frac{1}{5}} = 1.522 \: \rm bits.
\end{equation*}

\begin{figure}[h] 
    \centering
    \includegraphics[width=0.4\textwidth]{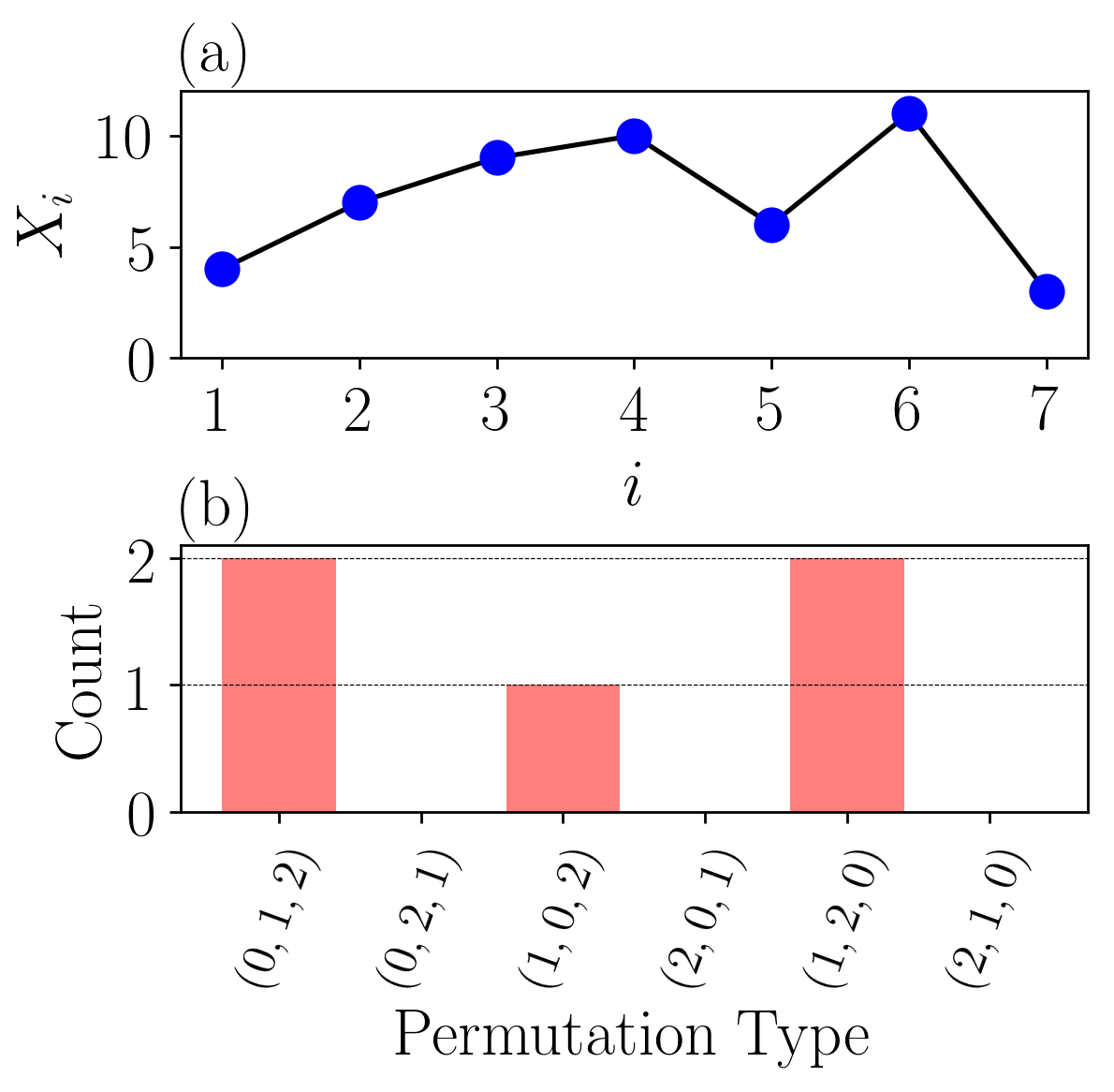}
    \caption{Abundance of each permutation from example data set $X$.}
    \label{fig: PE_Motif_Bar_Example}
\end{figure}

The permutation distribution can be visually understood by illustrating the probabilities of each permutation as separate bins. To accomplish this, Fig.~\ref{fig: PE_Motif_Bar_Example} was created by taking the same series $X$ (Fig.~\ref{fig: PE_Motif_Bar_Example}a) and placing the abundance of each permutation into its respective bin (Fig.~\ref{fig: PE_Motif_Bar_Example}b).
PE is at a maximum when all $n!$ possible permutations are evenly distributed or, equivalently, when the permutations are equiprobable with $p(\pi_i) =  p(\pi_{1}),  p(\pi_{2}),\ldots, p(\pi_{n!}) = \frac{1}{n!}$. From this, the maximum permutation entropy $H_{\rm max}$ is quantified as
\begin{equation} 
\begin{split}
H_{\rm max}(n) &= -\sum{p(\pi_i) \log{p(\pi_i)}} \\
 &= -\log{\frac{1}{n!}} = \log{n!}.
\end{split}
\label{eq: PEmax}
\end{equation}
Applying Eq.~\eqref{eq: PEmax} for $n=3$ yields a maximum PE of approximately 2.585 bits. Using the maximum possible entropy $\log_2{n!}$, the normalized permutation entropy is calculated as
\begin{equation} 
h_n  = -\frac{1}{\log_2{n!}} \sum{} p(\pi_i) \log_2{p(\pi_i)}.
\label{eq: PEn}
\end{equation}
Applying Eq.~\eqref{eq: PEn} to the example series $X$ results in $h_3 \approx 0.5888$.

\section{An overview of tools from TDA}
\label{sec:tda_overview}
Our TDA-based approaches for finding the delay dimension employs two types of persistence applied to two different types of data. 
Specifically, in our approach we combine the $0$-D sublevel persistence of one-dimensional time series with the $z$-score. 
This section provides a basic background of the topics needed to intuitively understand the subsequent analysis. 
More specifics can be found in \cite{Ghrist2008,Carlsson2009,Edelsbrunner2009,Oudot2015}.

\subsection{Simplicial complexes}
\label{sec:SimplicialComplexes}
An abstract $k$-simplex $\sigma$ is defined as a set of $k+1$ indices where $\dim(\sigma)=k$. 
If we apply a geometric interpretation to a $k$-simplex, we can think of it as a set $V$ of $k+1$ vertices. 
Using this interpretation, a $0$-simplex is a point, a $1$-simplex is an edge, a $2$-simplex is a triangle, and higher dimensional versions can be similarly obtained. 

A simplicial complex $K$ is a set of simplices $\sigma \subseteq V$ such that for every $\sigma \in K$, all the faces of $\sigma$, i.e., all the lower dimensional component simplices $\sigma' \subset \sigma$ are also in $K$. 
For example, if a triangle ($2$-simplex) is in a simplicial complex $K$, then so are the edges of the triangle ($1$-simplices) as well as all the nodes in the triangle ($0$-simplices). 
The dimension of the resulting simplicial complex is given by the largest dimension of its simplices according to $\dim(K) = \max _{\sigma \in K} \dim(\sigma)$.
The $n$-skeleton of a simplicial complex $K^{(n)}$ is the restriction of the latter to its simplices of degree at most $n$, i.e., $K^{(n)} = \{ \sigma \in K \mid \dim(\sigma) \leq n\}$.

Given an undirected graph $G = (V,E)$ where $V$ are the vertices and $E$ are the edges, we can construct the clique (or flag) complex
\begin{equation*}
 K(G) = \{ \sigma \subseteq V \mid uv \in E \text{ for all } u\neq v \in \sigma\}.
\end{equation*}

\subsection{Homology}
If we fix a simplicial complex $K$, then homology groups can be used to quantify the 1-dimensional topological features of the structure in different dimensions. 
For example, in dimension $0$, the rank of the $0$ dimensional homology group $H_0(K)$ is the number of connected components. 
The rank of the $1$-dimensional homology group $H_1(K)$ is the number of loops, while the rank of $H_2(K)$ is the number of voids, and so on. 
The homology groups are constructed using linear transformations termed boundary operators. 

To describe the boundary operators, we let $\{\alpha_{\sigma} \}$ be coefficients in a field $F$ (in this paper we choose $F =  \mathbb{Z}_2$). Since we are using the field $\mathbb{Z}_2$ we do not need to consider the orientation of the simplicial complex. 
Then $K^{(n)}$, the $n$-skeleton of $K$, can be used as a generating set of the $F$-vector space $\Delta_n(K)$. 
In this representation, any element of $\Delta_n(K)$ can be written as a finite formal sum $\sum\limits_{\sigma \in K^{(n)}}{\alpha_{\sigma} \sigma}$ called an $n$-chain. 
Further, elements in $\Delta_n(K)$ are added by adding their coefficients. 
The group of all $n$-chains is the $n$th chain group $\Delta_n(K)$, which is a vector space. Given a simplicial complex $K$, the boundary map $\partial_n: \Delta_n(K) \to \Delta_{n-1}(K)$ is defined by
\begin{equation*}
\partial_n([v_0, \ldots, v_n]) = \sum\limits_{i=0}^n{(-1)^i [v_0, \ldots, \hat{v}_i, \ldots, v_n]},
\end{equation*}
where $\hat{v}_i$ denotes the absence of element $v_i$ from the set. 
This linear transformation maps any $n$-simplex to the sum of its codimension $1$ (codim-$1$) faces. 
The geometric interpretation of the boundary operator is that it yields the orientation-preserved boundary of a chain.

By combining boundary operators, we obtain the chain complex
\begin{equation*}
\ldots \xrightarrow{\partial_{n+1}} \Delta_n(K) \xrightarrow{\partial_n} \ldots \xrightarrow{\partial_1} \Delta_1(K) \xrightarrow{\partial_0} 0,
\end{equation*}
where the composition of any two subsequent boundary operators is zero, i.e., $\partial_{n} \circ \partial_{n+1}=0$. 
An $n$-chain $\alpha \in \Delta_n(K)$ is a cycle if $\partial_n(\alpha) = 0$; it is a boundary if there is an $n+1$-chain $\beta$ such that $\partial_{n+1}(\beta) = \alpha$.
Define the kernel of the boundary map $\partial_n$ using $Z_n(K)=\{c \in \Delta_n(K): \partial_n c = 0\}$, and the image of $\partial_{n+1}$ $B_n(K)=\{c \in \Delta_n(K): c=\partial_{n+1} c', c' \in \Delta_{n+1}(K)\}$. 
Consequently, we have $B_k(K) \subseteq Z_k(K)$. 
Therefore, we define the $n$th homology group of $K$ as the quotient group $H_n(K)=Z_n(K)/B_n(K)$. 
In this paper, we only need $0$-dimensional persistent homology, and we always assume homology with $\mathbb{Z}_2$ coefficients which removes the need to keep track of orientation.
In the case of $0$-dimensional homology, there is a unique class in $H_0(K)$ for each connected component of $K$.
For $1$-dimensional homology, there is one homology class in $H_1(K)$ for each \textit{hole} in the complex and so on for higher dimensional invariants.

\subsection{Filtration of a simplicial complex}
\label{sec:FiltSimplicialComplexes}
Now, we are interested in studying the structure of a changing simplicial complex. 
We introduce a real-valued filtration function on the simplicies of $K$ such that $f(\tau)\leq f(\sigma)$ for all $\tau \leq \sigma$ simplices in $K$. 
If we let $\{y_1 < y_2 < \ldots < y_{\ell}\}$ be the set of the sorted range of $f$ for any $y \in \mathbb{R}$, then the filtration of $K$ with respect to $f$ is the ordered sequences of its subcomplexes
\begin{equation*}
  \emptyset \subseteq K(y_1) \subseteq K(y_2) \subseteq \cdots \subseteq K(y_{\ell})=K.
\end{equation*}
The sublevel set of $K$ corresponding to $y$ is defined as 
\begin{equation}
\label{eq:filtration}
	K(y) = \{\sigma \in K \mid f(\sigma) \leq y\},
\end{equation}
where each of the resulting $K(y)$ is a simplicial complex, and for any $y_1 \leq y_2$, we have $K(y_1) \subseteq K(y_2)$. 

The filtration of $K$ enables the investigation of the topological space under multiple scales of the output value of the filtration function $f$. 
In this paper we consider a filtration function which corresponds to 0D sublevel persistence applied to $1$-D time series data.

\paragraph{$0$-D persistence applied to $1$-D time series:}
Let $\chi$ be the time-ordered set of the critical values of a time series. 
Here, we can think of the simplicial complex $K=G(V,E)$ containing a number of vertices $|V|$ equal to the number of critical values in the time series and only the edges $E$ that connect adjacent vertices. i.e., vertices $\{v_i \mid 1\leq i \leq n\}$, and edges $\{v_i v_{i+1} \mid 1\leq i \leq n-1 \}$.
Therefore, we have a one-to-one correspondence between the critical values in $\chi$ and the vertices of the simplicial complex $V$.
We define the filtration function for every face $\sigma$ in $K$ according to
\begin{equation*}
f(\sigma_i) = \begin{cases}
	\chi_i & \text{ if }\sigma_i \in V, \\
	\max{(u,v)}      & \text{ if }\sigma_i=uv \in E.
	\end{cases}
\end{equation*}
Using this filtration function in Eq.~\eqref{eq:filtration}, we can define an ordered sequence of subcomplexes where $y\in[\min{(\chi)}, \max{(\chi)}]$.

\subsection{Persistent homology} \label{ssec:persistent_homology}
Persistent homology is a tool from topological data analysis which can be used to quantify the shape of data.
The main idea behind persistent homology is to watch how the homology changes over the course of a given filtration.

Fix a dimension $n$, then for a given filtration
\begin{equation*}
  K_1 \subseteq K_2 \subseteq \cdots \subseteq K_N
\end{equation*}
we have a sequence of maps on the homology
\begin{equation*}
  H_n(K_1) \to H_n(K_2) \to \cdots \to H_n(K_N).
\end{equation*}
We say that a class $[\alpha] \in H_n(K_i)$ is born at $i$ if it is not in the image of the map $H_n(K_{i-1}) \to H_n(K_i)$.
The same class dies at $j$ if $[\alpha] \neq 0$ in $H_n(K_{j-1})$ but $[\alpha] = 0$ in $H_n(K_{j})$.

\begin{figure}[h] 
    \centering
    \includegraphics[width=0.45\textwidth]{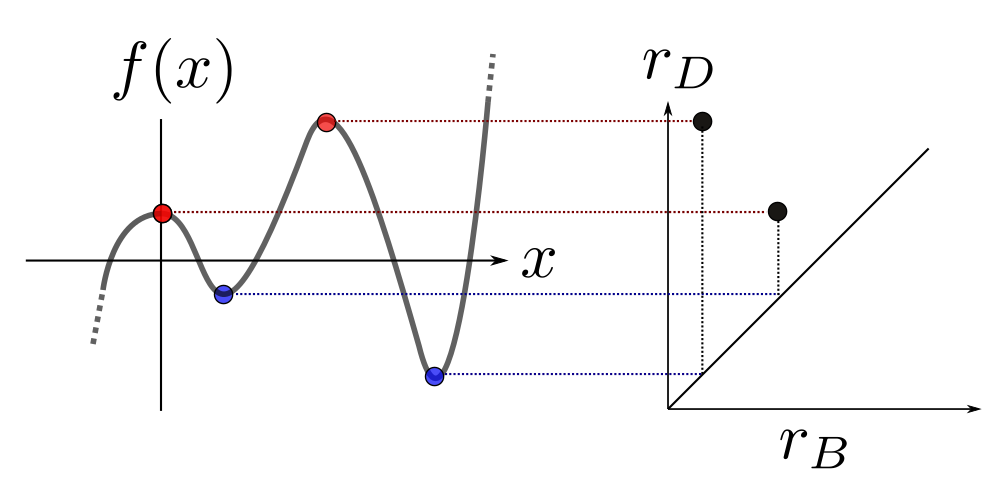}
    \caption{Example formulation of a persistence diagram based on $0$-D sublevel sets.}
    \label{fig:persexample}
\end{figure}

This information can be used to construct a persistence diagram as follows. 
A class that is born at $i$ and dies at $j$ is represented by a point in $\mathbb{R}^2$ at $(i,j)$. 
The collection of the points in the persistence diagram, therefore, give a summary of the topological features that persists over the defined filtration. 
See the example of Fig.~\ref{fig:persexample} for $n=0$ and time series data.

\section{Permutation Delay} \label{sec:timeLag}
To form permutations from a time series a delay embedding is applied to a uniform subsampling of the original time series according to the embedding parameter $\tau$. 
For example, the subsampled sequence $X$ with elements $\{x_i: i \in \mathbb{N}\cup 0 \}$ subject to the delay $\tau$ is defined as $X(\tau) = [x_0, x_{\tau}, x_{2\tau}, \ldots]$. 
Riedl et al.~\cite{Riedl2013} showed that PE is sensitive to the time delay, which prompts the need for a robust method for determining an appropriate value for $\tau$. 
For estimating the optimal $\tau$, we will be investigating the following methods in the subsequent sections: Mutual Information (MI) in Section~\ref{sec:MI}, and combining $0$-D persistence with the frequency and time domains (Section \ref{ssec:0d_sublevel_for_delay}). 
We recognize, but do not investigate, some other commonly used methods for finding $\tau$. 
These include the autocorrelation function~\cite{grassberger1983measuring} and the phase space expansion~\cite{buzug1992optimal}.

Before introducing the TDA-based methods, in Section~\ref{ssec:relating_PE_to_TDA} we elucidate a connection between the permutation delay parameter and the state space reconstruction delay parameter for Takens' embedding to justify the use of methods such as MI and the TDA-based methods.

\subsection{Mutual Information} \label{sec:MI}
A common method for selecting the delay $\tau$ for state space reconstruction is through a measure of the mutual information between a time series $x(t)$ and its delayed version $x(t+\tau)$~\cite{Fraser1986}.
Mutual information is a measurement of how much information is shared between two sequences, and it was first realized by Shannon et al.~\cite{Shannon1951} as 
\begin{equation} 
I(X;Y) = \sum_{x \in X} \sum_{y \in Y}p(x,y)\log\frac{p(x,y)}{p(x)p(y)},
\label{eq: MI}
\end{equation}
where $X$ and $Y$ are separate sequences, $p(x)$ and $p(y)$ are the probability of the element $x$ and $y$ separately, and $p(x, y)$ is the joint probability of $x$ and $y$.  Fraser and Swinney~\cite{Fraser1986} showed that for a chaotic time series the mutual information between the sequence $x(t)$ and $x(t+\tau)$ will decrease as $\tau$ increases until reaching a minimum.
At this delay $\tau$, the individual data points share minimum amount of information, thus indicating that the data points are sufficiently separated.
While this delay value was specifically developed for phase space reconstruction from a single time series, we show in Section~\ref{ssec:relating_PE_to_TDA} how this delay parameter selection method is also appropriate for permutation entropy. Due to this relationship and the result in \cite{Myers2020} suggesting the use of MI for PE delay selection, we benchmark the delays $\tau$ obtained from our methods against their MI counterparts.

\subsection{Relating Permutations to Delay Reconstruction} \label{ssec:relating_PE_to_TDA}
The goal in this section is to relate the distribution of permutations formed from a given delay $\tau$ to the state space reconstruction with the same delay $\tau$. This connection will show the time delay for both permutations and state space reconstruction are related. 

\begin{figure}[h!] 
    \centering
    \includegraphics[scale = 0.49]{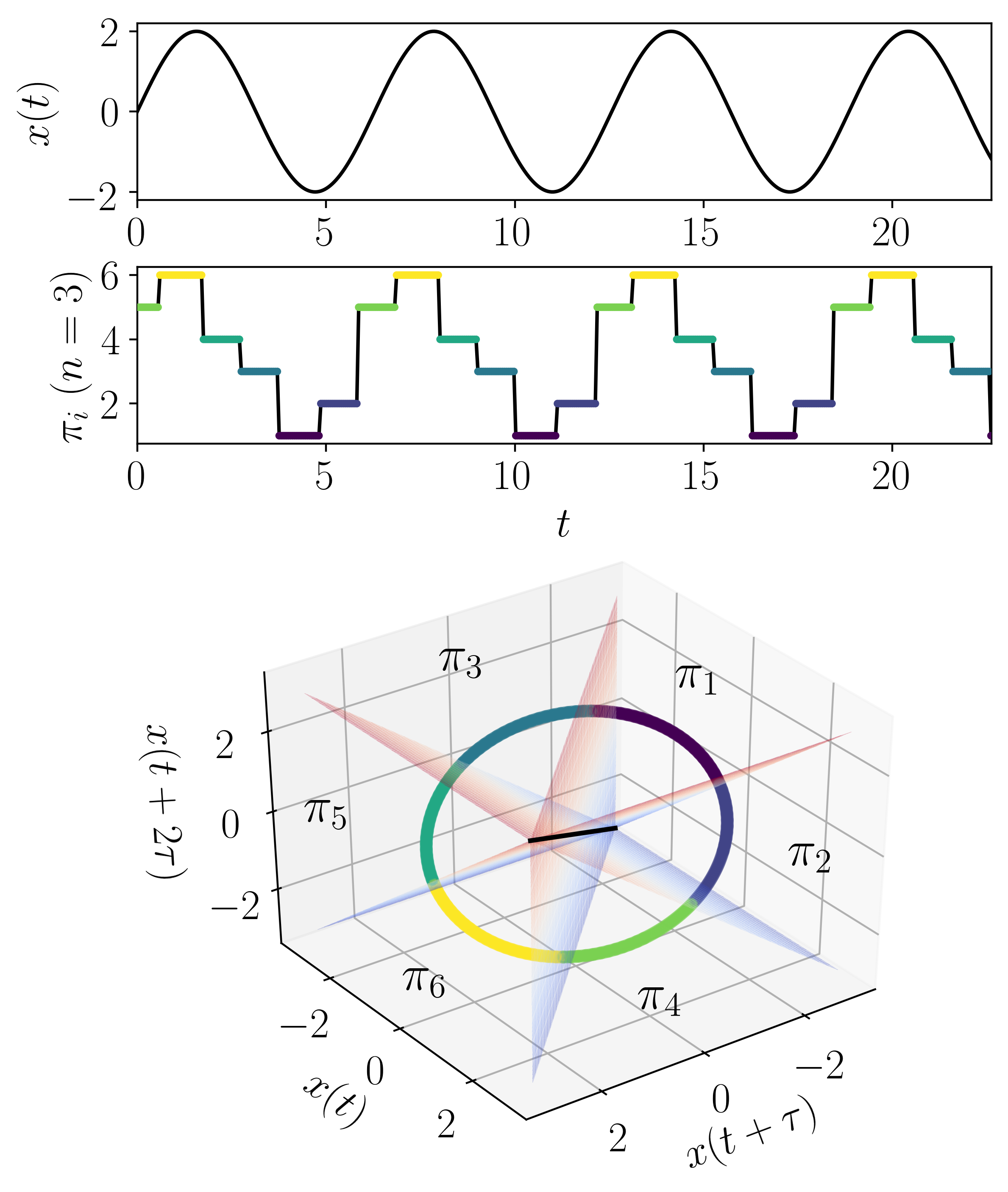}
    \caption{Example formation of a permutation sequence from the time series $x(t) = 2\sin(t)$ with sampling frequency $f_s = 20$ Hz, permutation dimension $n=3$ and delay $\tau = 40$. The corresponding time-delay embedded vectors from $x(t)$ with the permutation binnings ($\pi_1, \ldots, \pi_6$) in the state space are shown in the bottom figure.}
    \label{fig:takens_to_perm_to_binning}
\end{figure}

\begin{figure*}[!b] 
    \centering
    \includegraphics[width=\textwidth]{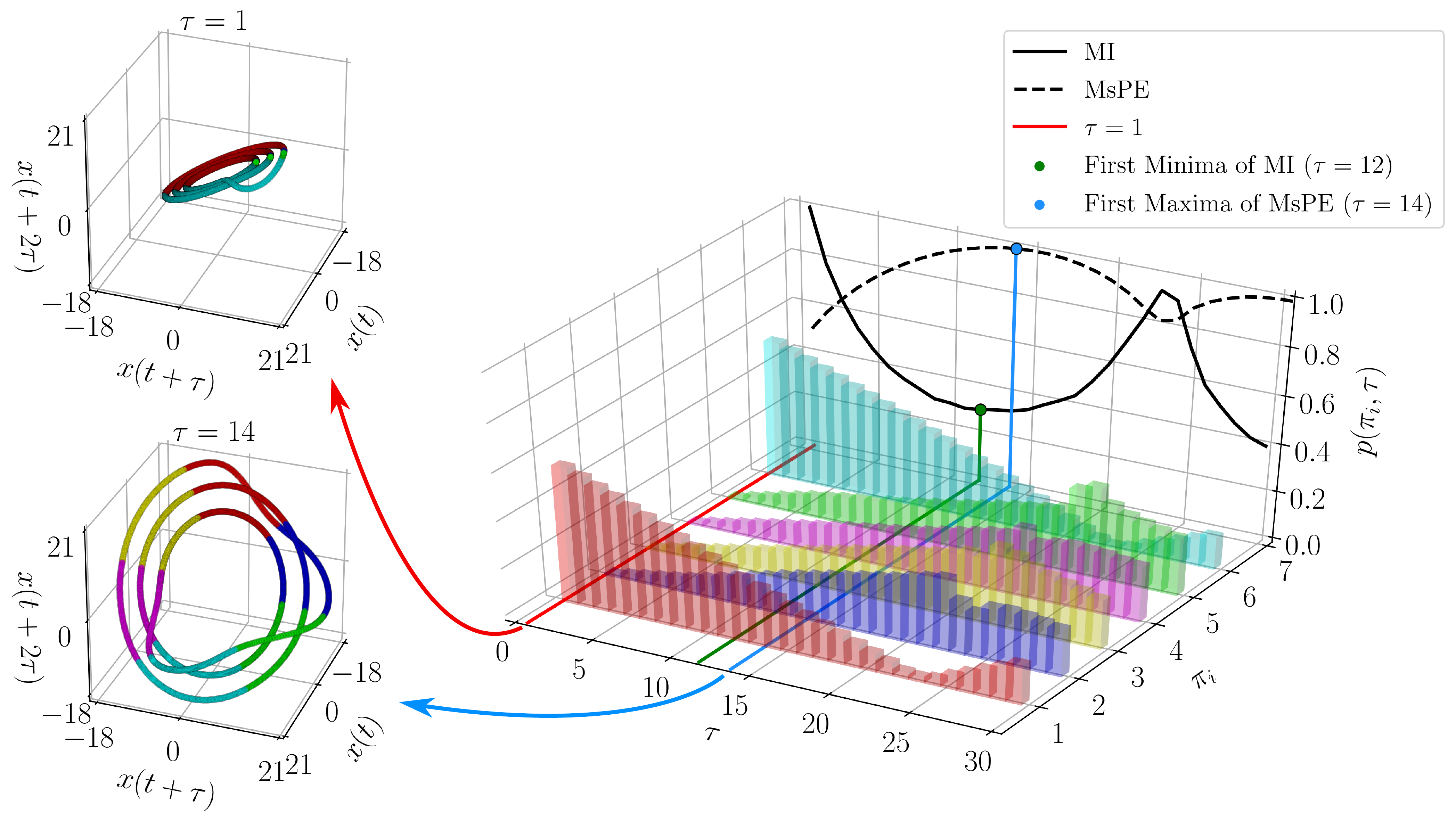}
    \caption{Example comparing first minima of mutual information and first maxima of multi-scale permutation entropy, which demonstrates the correspondence between the two. On the left are the $n=3$ time delayed state space reconstructions with an inaccurately chosen $\tau = 1$ and appropriate $\tau = 14$. On the right shows the permutation distribution as $\tau$ increases and the associated multi-scale permutation entropy and mutual information plots.}
    \label{fig:relating_perms_to_takens_horizontal}
\end{figure*}

Let us first start by describing the process for state space reconstruction and its similarity to permutations. As described by Takens'~\cite{Takens1981}, we can reconstruct an attractor that is topologically equivalent to the original attractor of a dynamical system through embedding a 1-D signal into $\mathbb{R}^n$ by forming a cloud of delayed vectors as $v_i = [x(t_i), x(t_{i+\tau}), x(t_{i+2\tau}), \dots, x(t_{i+(n-1)\tau})]$ for $i \in [0,L-n\tau]$, where $L$ is the length of the discretely and uniformly sampled signal. 
Permutations are formed in a very similar fashion where we take our vectors $v_i$ and find their symbolic representation based on their ordinal ranking as explained in Section~\ref{sec:PE_example}. The different permutation types can be viewed as an inequality-based binning of the $\mathbb{R}^n$ vector space of the reconstructed dynamics as shown in Fig.~\ref{fig:takens_to_perm_to_binning} for dimension $n=3$. For example, consider the region in Fig.~\ref{fig:takens_to_perm_to_binning} labeled $\pi_1$. All points on the trajectory shown in this region have the property that  $x(t+2\tau)>x(t+\tau)>x(t)$ meaning that points in this region are placed into the group $\pi_1$ where all nodes in the permutation are increasing. Similarly, the other sections of the trajectory are appropriately binned to the remaining permutations. This provides a first intuitive understanding of the connection between permutation and state space reconstruction; however, we need to determine if the optimal $\tau$ parameter used in $v_i$ is related to the optimal delay in PE.

Takens' embedding theorem explains that, theoretically, any delay $\tau$ would be suitable for reconstructing the original topology of the attractor; however, this has the requirement of unrestricted signal length with no additive noise in the signal~\cite{Takens1981}. Since this is rarely a condition found in real-world signals, a $\tau$ is chosen to unfold the attractor such that noise has a minimal effect on the topology of the reconstructed dynamics.

Let us now explain what we mean by the correspondence between $\tau$ and the unfolding of the dynamics and what effect this has on the corresponding permutations. 
If the delay $\tau$ is too small (e.g. $\tau = 1$ for a continuous dynamical system with a high sampling rate) the delay embedded reconstructed attractor will be clustered around the hyper-diagonal in $\mathbb{R}^n$. Additionally, the corresponding permutations will be overwhelmingly dominated by the permutation types $\pi_1$ and $\pi_{n!}$ with these two permutations being of the all increasing and all decreasing ordinal patterns, respectively. The dominance of these two permutations for a delay $\tau$ that is too small was termed by Casdagli et al.~\cite{casdagli1991state} as the ``redundancy effect." For an example of this see the permutation distribution and clustering about the hyper-diagonal in $\mathbb{R}^3$ in Fig.~\ref{fig:relating_perms_to_takens_horizontal} when $\tau = 1$. We see that for $\tau = 1$, the trajectory is dominated by $\pi_1$ and $\pi_6$ permutations, and for this case each group makes up a probability of 50\% of the embedded attractor. As the delay increases, the probabilities of the remaining groups begin to increase and the $\pi_1$ and $\pi_6$ probabilities decrease. These probabilities are generally calculated by the computing the fraction of the trajectory that is occupied by each ordinal pattern. This example is based on the $x$-solution to the periodic Rossler dynamical system as described in Section~\ref{app:rossler}. As the delay increases beyond the redundancy effect, the reconstructed attractor begins to unfold to have a similar shape and topology as the true attractor. Correspondingly, as the delay increases the permutation distribution tends towards a more equiprobable distribution (See Fig.~\ref{fig:relating_perms_to_takens_horizontal} at $\tau \approx 14$). 

A way of summarizing the permutation probability distribution is actually through PE itself and more specifically the analysis of Multi-scale Permutation Entropy (MsPE). Riedl et al.~\cite{Riedl2013} showed how after the redundancy effect there is a suitable delay for PE, which we related to the first maxima of the MsPE plot~\cite{Myers2020}. The MsPE plot for our periodic Rossler example is shown in Fig.~\ref{fig:relating_perms_to_takens_horizontal}. Let us also look at the MI plot as a comparison. The idea behind MI is that at the first minima of the mutual information between $x(t_i)$ and $x(t_{i+\tau})$ the delay $\tau$ accurately provides a suitable delay for state space reconstruction. A quick investigation of the MI function reveals a high degree of correlation between the MI function and the MsPE function with the first maxima of MsPE being approximately at the same $\tau$ as the first minima of MI. 
When the delay becomes significantly larger than the first minima of MI or maxima of MsPE, the permutation distribution begins to fluctuate as shown in Fig.~\ref{fig:relating_perms_to_takens_horizontal}. This effect was termed the ``irrelevance effect" by Casdagli et al.~\cite{casdagli1991state}. This increasing of $\tau$ beyond the the first minima also correlates with what Kantz and Schreiber~\cite{Kantz2003} describe as the reconstruction filling an overly large space with the vectors already being independent. Additionally, at a minima beyond the first minima, Fraser and Swinney~\cite{Fraser1986} showed how the reconstructed attractor shape will no longer qualitatively match the shape of the true state space.

In summary, we have shown a main heuristic result we need to move forward: tools for delay parameter selection for PE can be suitable for state space reconstruction and vice-versa. While we do not provide a proof that PE and state space reconstruction use the same $\tau$, it has recently been shown that a connection between co-homology, information theory, and probability does exist~\cite{Baudot2015}, which strengthens our qualitative analysis of this connection. In the following sections we leverage tools from TDA to determine the optimal $\tau$ associated with the unfolding of the attractor.

\subsection{Finding $\tau$ Using Sublevel Set Persistence} \label{ssec:0d_sublevel_for_delay}
In this section our goal will be to leverage sublevel set persistence for the selection of $\tau$ for both state space reconstruction and permutation entropy. 
Specifically, our goal is to automate the frequency analysis method~\cite{Melosik2016} for selecting $\tau$ for state space reconstruction by analyzing both the time and frequency domain of the signal using sublevel set persistence.
Melosik and Marszalek~\cite{Melosik2016} leveraged Shannon-Nyquist sampling criteria and used  the maximum significant frequency $f_{\rm max}$ and the sampling frequency $f_s$ to select an appropriate $\tau$ according to
\begin{equation}
	\tau = \frac{f_s}{\alpha f_{\rm max}},
	\label{eq:fs_fmax}
\end{equation}
where $\alpha \in [2, 4]$. The value $\alpha = 2$ is associated to the Nyquist sampling rate, while $\alpha > 4$ produces an oversampling. Since this method was developed using the Nyquist sampling rate, it is applicable for continuous, band-limited signals. This frequency based approach was used to find suitable delays for the 0/1 test on chaos and heuristically compare the Lorenz attractor and its time-delay reconstruction~\cite{Melosik2016}. The heuristic comparison showed that this frequency approach actually provided more accurate delay parameter selections for state space reconstruction than the mutual information function when trying to replicate the shape of the attractor.
Unfortunately, a major drawback of this method is the non-trivial selection of $f_{\max}$. In Melosik's and Marszalek's original work~\cite{Melosik2016} the maximum frequency was manually selected using a normalized Fast Fourier Transform (FFT) cutoff of approximately 0.01, which does not address the possibility of additive noise.

In~\cite{Myers2020} the maximum ``significant'' frequency was approximated in a time series using the FFT and defining a power spectrum cutoff based on the statistics of additive noise in the FFT. An issue with this method for non-linear time series is that the Fourier spectrum does not easily yield itself to selecting the maximum ``significant'' frequency for chaotic time series even with an appropriately selected cutoff to ignore additive noise. Additionally, the method was only developed for Gaussian White Noise (GWN) contamination of the original time series.

The following sections improve the selection of the maximum significant frequency using two novel methods based on 0-D sublevel set persistence. We chose to use 0-D sublevel set persistence due to its computational efficiency and stability for true peak selection~\cite{Khasawneh2018, Cohen-Steiner2006}. The first method is based on a time domain analysis of the sublevel set lifetimes (see Section~\ref{sssec:max_freq_time_domain}), and the second implements a frequency domain analysis using sublevel set persistence and the modified $z$-score (see Section~\ref{sssec:max_freq_frequency_domain}). 

\subsubsection{Time Domain Approach} \label{sssec:max_freq_time_domain}

\begin{figure}[h] 
    \centering
    \includegraphics[width=0.5\textwidth]{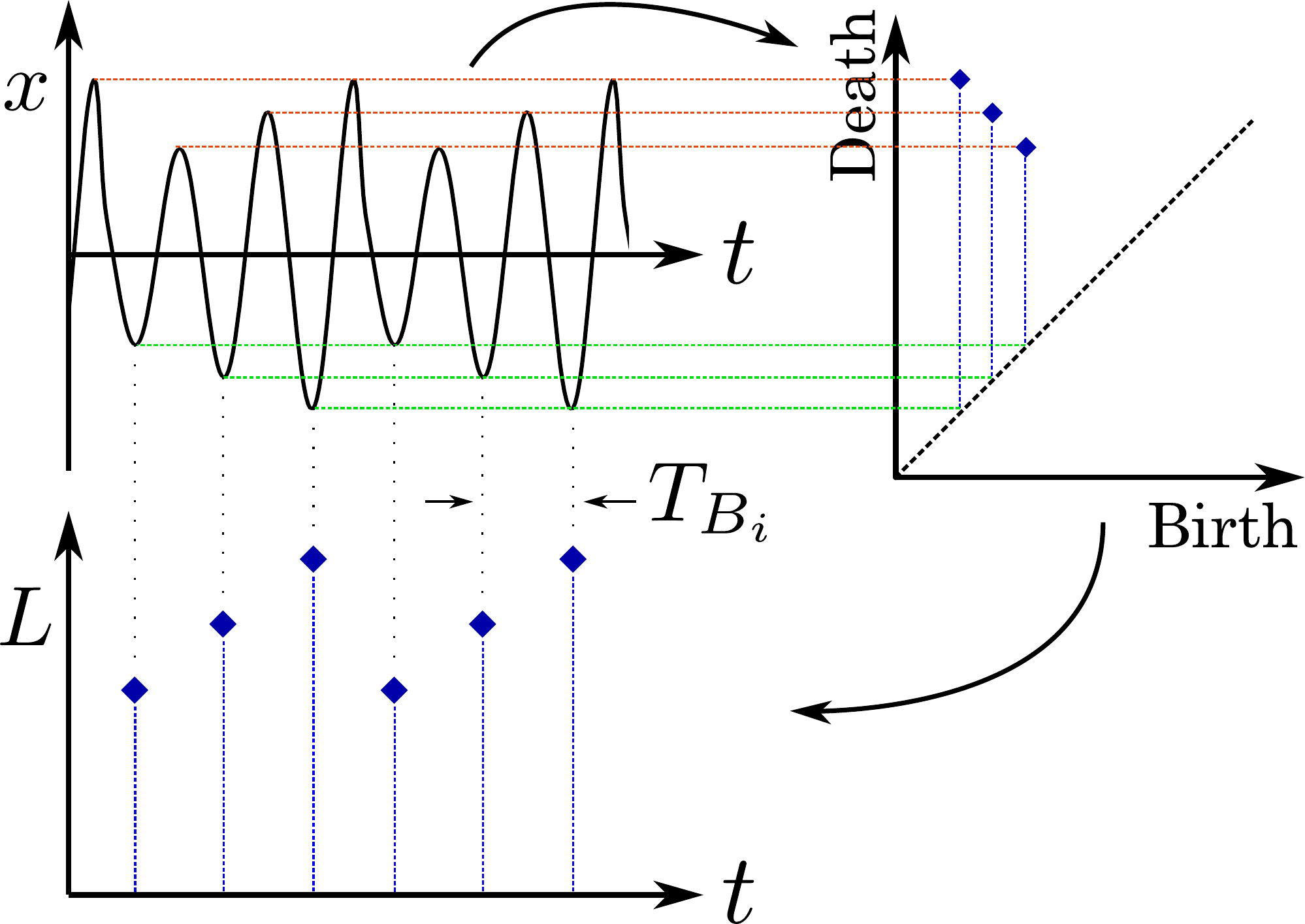}
    \caption{Example demonstrating process from time series $x$ (periodic Rossler system) to sublevel set persistence diagram to time ordered lifetimes on the bottom left. Additionally, on the bottom left shows a sample time periodic between sublevel sets as $T_{B_i}$.}
    \label{fig:TS_to_timeOrderedLifetimes_with_T_B_i}
\end{figure}

The first approach we implement for estimating the maximum significant frequency of a signal is based on a time domain analysis of the sublevel set persistence.
This process uses the time ordered lifetimes from the sublevel set persistence diagram. Time ordered lifetimes and a cutoff separating the sublevel sets associated with noise were previously introduced in~\cite{Myers2020a}. Here we use those methods and results to find the time $t_B$ in which all the significant sublevel sets are born. Figure~\ref{fig:TS_to_timeOrderedLifetimes_with_T_B_i} shows an example time order lifetimes plot where the time between two adjacent lifetimes is defined as $T_{B_i}$. 
If we use $T_{B_i}$ as an approximation of a period in the time series, then we can calculate the associated frequencies as $f_i = 1/T_{B_i}$ Hz. If we then look at the distribution of $f_i$, the maximum ``significant" frequency can be approximated using the 75\% quantile of the distribution of the frequencies as $f_{\rm max} \approx Q_{75}(f)$. This quantile allows for a few outlier frequencies to occur without having a significant effect on the estimate of the maximum significant frequency.

\begin{figure}[h] 
    \centering
    \includegraphics[width=\textwidth]{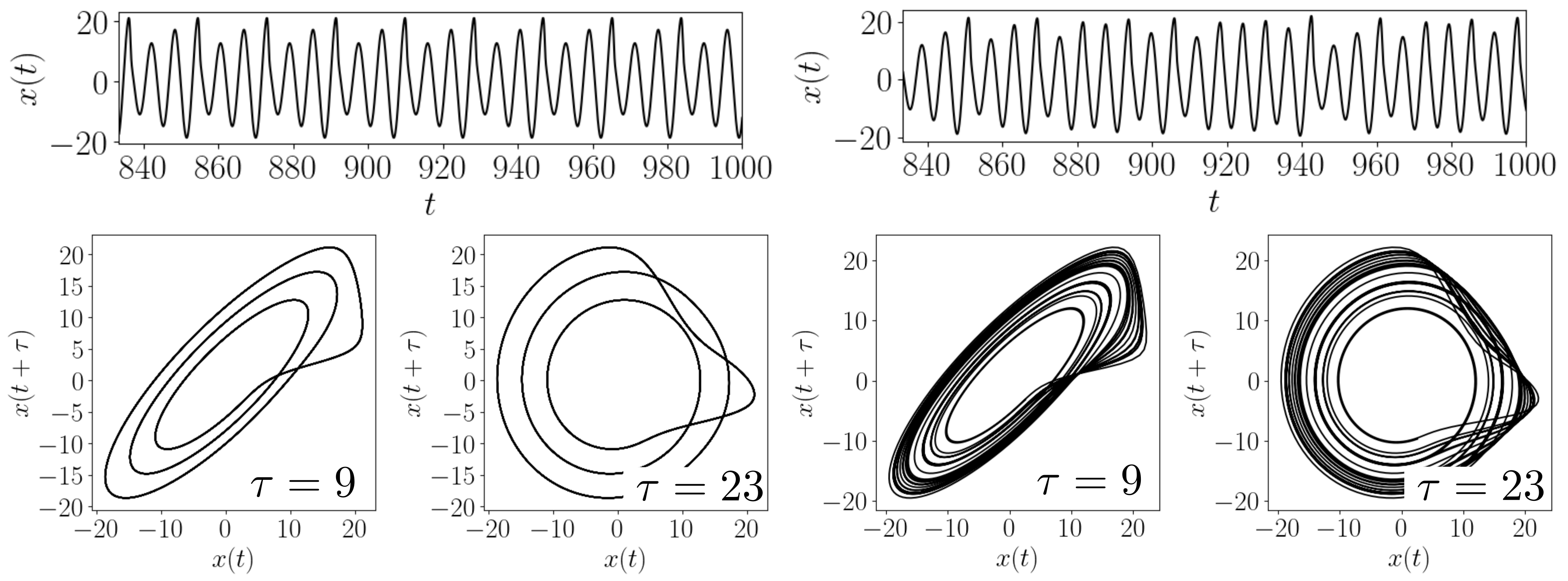}
    \caption{Example demonstrating the time delay $\tau = 23$ result for the periodic and chaotic Rossler example time series shown in the top figures and the resulting $n=2$ Takens' embeddings. The bottom right figures show the embeddings using the delays from \cite{Tao2018} ($\tau=9$) and the sublevel persistence method ($\tau=23$) for both periodic (left) and chaotic (right) signals.}
    \label{fig:periodic_rossler_to_SL_for_delay}
\end{figure}

Applying this method to the periodic Rossler system described in Eq.~\eqref{eq:rossler} results in $\tau = 23$ with the corresponding state space reconstruction for $n=2$ shown in Fig.~\ref{fig:periodic_rossler_to_SL_for_delay}. This suggested delay is larger than that of mutual information ($\tau = 16$), but relatively close. We also see in Fig.~\ref{fig:periodic_rossler_to_SL_for_delay} that the attractor appears to be circular. Embedding this signal using the suggested delay from \cite{Tao2018} resulted in an attractor that was more elliptical in shape which is not ideal for attractor reconstruction. This suggests that the time-domain analysis for selecting the maximum frequency and corresponding delay functions can automatically suggest an appropriate delay for permutation entropy and state space reconstruction resulting in a more optimal unfolding of the attractor in both periodic and chaotic cases.

\begin{figure*}[h] 
    \centering
    \includegraphics[width=\textwidth]{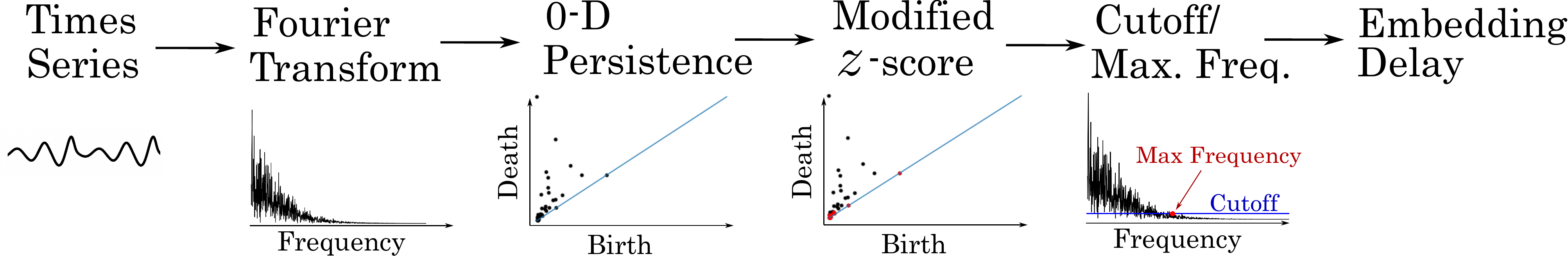}
    \caption{Overview of procedure for finding maximum significant frequency using $0$-dimensional sublevel set persistence and the modified $z$-score for a signal contaminated with noise.}
    \label{fig: 0d_procedure}
\end{figure*}
\subsubsection{Fourier Spectrum Approach} \label{sssec:max_freq_frequency_domain}

In this section we present a novel TDA based approach for finding the noise floor in the Fourier spectrum for selecting the maximum significant frequency $f_{\max}$ to be used for selecting $\tau$ for PE through Eq.~\eqref{eq:fs_fmax}. Specifically, we show how the $0$-dimensional sublevel set persistence, a tool from TDA discussed in Section~\ref{sec:tda_overview}, can be used to find the significant lifetimes and the associated frequencies in the frequency spectrum. 
Although it would be ideal to separate the significant lifetimes based on propagating the FFT of a random process into the persistence space, this task is not trivial.
There have been studies on pushing forward probability distributions into the persistence domain~\cite{adler2010persistent,Adler2014,Kahle2013}, but it is difficult to obtain a theoretical cutoff value in persistence space; therefore, we instead separate the noise lifetimes from significant lifetimes through the use of the modified $z$-score.
This separation allows us to find the noise floor and the maximum significant frequency via a cutoff. This process for finding the cutoff and associated maximum frequency is illustrated in Fig.~\ref{fig: 0d_procedure}. The following paragraphs give an overview of the modified $z$-score and cutoff analysis.

\paragraph{Modified $z$-score}

The modified $z$-score $z_m$ is essential to understanding the techniques used for isolating noise from a signal~\cite{seo2006review}. The standard score, commonly known as the $z$-score, uses the mean and the standard deviation of a data set to find an associated $z$-score for each data point and is defined as
\begin{equation} 
z = \frac{x - \mu}{\sigma}, 
 \label{eq: zscore}
\end{equation}
where $x$ is a data point, $\mu$ is the mean, and $\sigma$ is the standard deviation of the data set, respectively. The $z$-score value is commonly used to identify outliers in the data set by rejecting points that are above a set threshold, which is set in terms of how many standard deviations away from the mean are acceptable. Unfortunately, the $z$-score is susceptible to outliers itself with both the mean and the standard deviation not being robust against outliers~\cite{leys2013detecting}. This led Hampel~\cite{hampel1974influence} to develop the modified $z$-score as an outlier detection method that is robust to outliers. The logic behind the modified $z$-score or median absolute deviation (MAD) method is grounded on the use of the median instead of the mean. The MAD is calculated as 
\begin{equation} 
{\rm MAD} = {\rm median}(|\mathbf{x} - \tilde{x}|), 
 \label{eq: MAD}
\end{equation}
where $\mathbf{x}$ is an array of data points and $\tilde{x}$ is its median. The MAD is substituted for the standard deviation in Eq.~\eqref{eq: zscore}. To improve the modified $z$-score, Iglewicz and Hoaglin~\cite{iglewicz1993volume} suggested to additionally substitute the mean with the median. The resulting equation for the modified $z$-score is then
\begin{equation} 
z_m = 0.6745\frac{\mathbf{x} - \tilde{x}}{\rm MAD},
 \label{eq: modfiedzscore}
\end{equation}
where the value 0.6745 was suggested in~\cite{iglewicz1993volume}.
We can now use the modified $z$-score $z_m$ for evaluating the ``significance" of each point in the sublevel set persistence diagram of the Fourier spectrum. A threshold for separating noise in the persistence domain is discussed in the following paragraph.

\begin{figure}[h] 
    \centering
    \includegraphics[width=0.6\textwidth]{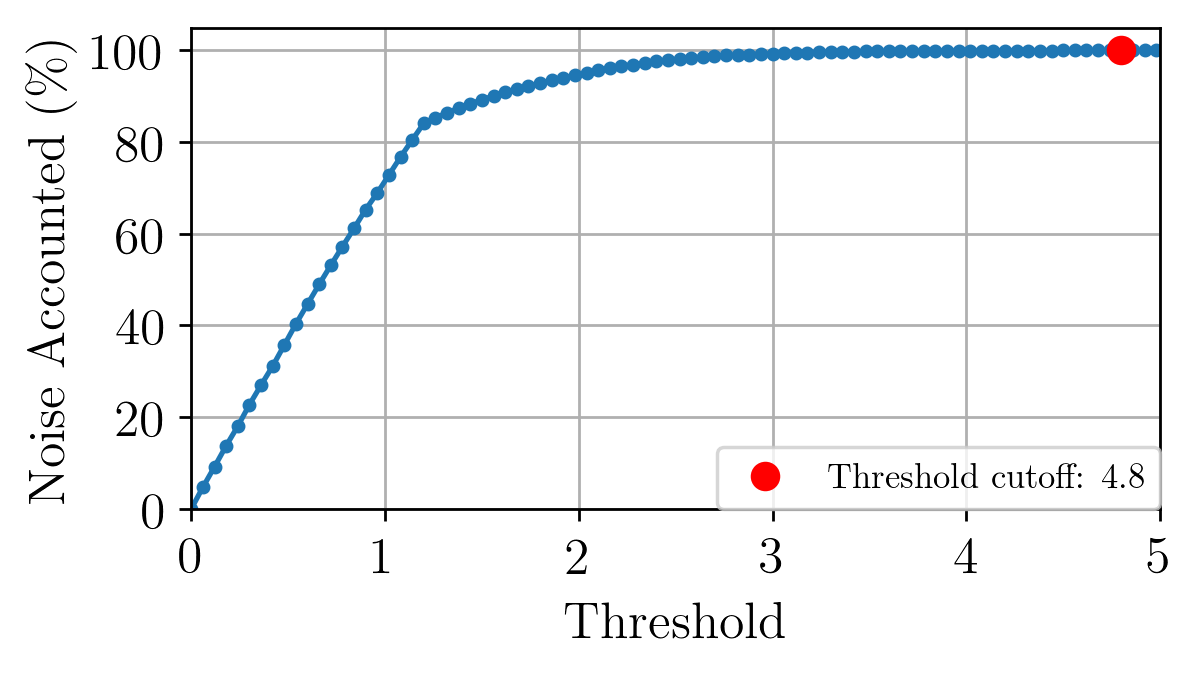}
    \caption{Percent of the persistence points from $0$-D sublevel set persistence of the FFT of GWN using the modified $z$-score with the provided threshold ranging from 0 to 5.}
    \label{fig:threshold_cutoff}
\end{figure}

\paragraph{Threshold and Cutoff Analysis}\label{persAnalysis}
To determine the noise floor in the normalized Fast Fourier Transform (FFT) spectrum, we compute the $0$-dimensional persistence of the FFT. This provides relatively short lifetimes for the noise, while the prominent peaks, which represent the actual signal, have comparatively long lifetimes or high persistence. To separate the noise from the outliers we calculate the modified $z$-score for the lifetimes in the persistence diagram. We can then determine if the lifetime is associated to noise or signal based on a $z_m$ cutoff as $D$, where we can label a lifetime as significant (an outlier) if $z_m > D$. Iglewicz and Hoaglin~\cite{iglewicz1993volume} suggest a $z_m$ threshold of $D = 3.5$ based on an analysis of 10,000 random-normal observations. However, we apply both the FFT and 0-D sublevel set persistence to the original signal so we need to determine if this cutoff is also suitable for data that has been processed in this way. To do this we used a signal of 10,000 random-normal observations and applied FFT. We then calculated the 0-D sublevel set persistence and computed the modified $z$-score $z_m$ using the resulting lifetimes. For an accurate cutoff we would expect to label all of the lifetimes as noise with $z_m < D$ since each signal is composed of pure noise. As shown in Fig.~\ref{fig:threshold_cutoff}, a threshold of approximately $D = 4.8$ labels all of the lifetimes as noise. This threshold was rounded up to 5 for simplicity. We can now simply define a cutoff based on the labeling of each lifetime from the modified $z$-score with $\rm Cutoff = \max (lifetime_{noise})$. We emphasize that this threshold is only appropriate for additive Gaussian white noise and a different cutoff may need to be obtained with a different noise distribution.

We can now find the maximum significant frequency $f_{\rm max}$ as the highest frequency in the Fourier spectrum with an amplitude greater than the specified cutoff. For this method to accurately function, it is required that there is some additive noise in the time series. This is due to the fact that our pipeline in Fig.~\ref{fig: 0d_procedure} does not distinguish between signals with and without noise so if a noise-free signal is used, the outliers will drastically alter the results and the cutoff will not be accurate. Adding noise to the signal is meant to increase the contrast between outliers and the true signal to allow for a proper cutoff to be obtained. Further, it is uncommon to have a signal without noise in real world situations. To accommodate this for our simulations, additive Gaussian noise with Signal-to-Noise Ratio of 30 dB is added to the time series before calculating the FFT. If we apply this method to the example periodic Rossler system time series we find a suggested delay of $\tau = 15$. Which is close to the delay estimated using mutual information. W will further investigate its accuracy on several other systems in Section~\ref{sec:results} to make more general conclusions on the functionality of this method for selecting $\tau$.

\section{Permutation Dimension} \label{sec:dimension}
In this section we will show that, contrary to the delay selection, the dimension for permutation entropy is not related to that of Takens' embedding. Additionally, we will provide a simple method for selecting an appropriate permutation dimension based on the permutation distribution. 

Permutation entropy is often used to differentiate between the complexity of a time series when there is a dynamic state change (e.g. periodic compared to chaotic), so the dimension should be chosen such that it is large enough to capture these changes. To accomplish this we suggest that permutations of the time series do not occupy all of the possible permutations, but rather only a fraction of the permutations when an appropriate delay is selected. This criteria is set so that a change can be captured by an increase/decrease in the number of permutations and their associated probabilities. Because of this, we suggest a dimension where, at most, only 50\% of the permutations are used. However, it may be more suitable to select a dimension where a lower percentage is used (e.g. 10\%).

To begin this method for determining if the dimension is high enough to capture the time series complexity we will define $N_\pi$ as the number of permutation types where the probability of that permutation type is significant. Specifically, we will consider the probability of that permutation to be significant if the number of occurrences of permutation $\pi$ is greater than 10\% of the maximum number of occurrences of any permutation type of dimension $n$. The permutation delay $\tau$ was selected from the expert suggested values provided in~\cite{Myers2020, Riedl2013}.
We can now express our needed dimension as the ratio and inequality
\begin{equation}
\frac{N_\pi}{n!} \leq R,
\label{eq:dimension_complexity_ratio}
\end{equation}
where $R = 0.50$ for the suggested maximum 50\% criteria.

\begin{figure}[h] 
    \centering
    \includegraphics[width=0.6\textwidth]{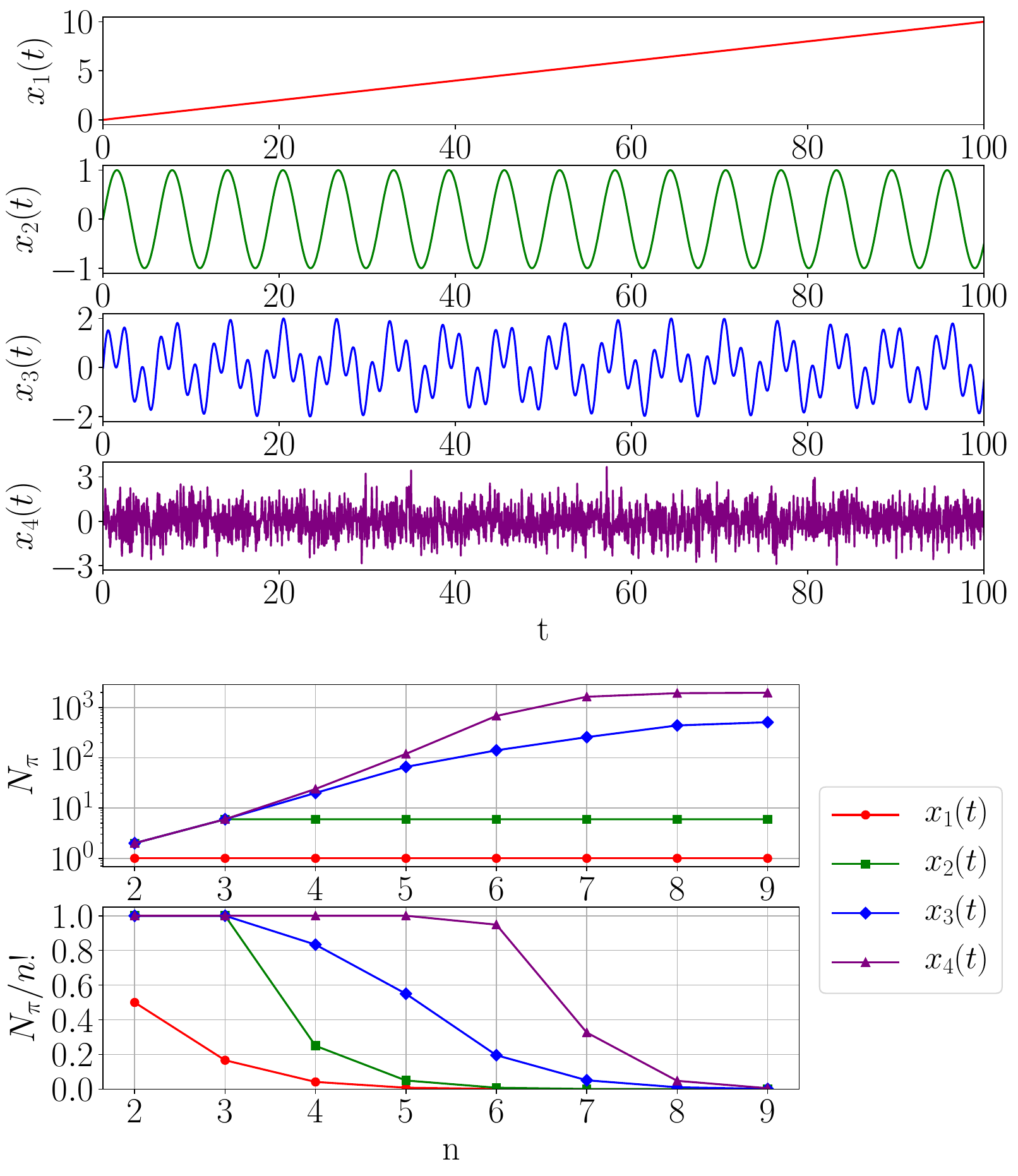}
    \caption{Percent of permutations used $R = N_\pi / n!$ for each example time series (see Eq.~\eqref{eq:trivial_examples}) as the dimension is incremented.}
    \label{fig:example_time_series_permutation_complexity_vs_dimension}
\end{figure}

To compare this dimension to the standard Takens' embedding tools for selecting $n$ we will implement four examples: 
\begin{equation}
\begin{split}
x_1(t) & = \frac{t}{10} \\
x_2(t) & = \sin(t) \\
x_3(t) & = \sin(t) + \sin(\pi t) \\
x_4(t) & \sim \mathcal{N}(\mu = 0, \sigma^2 = 1),
\end{split}
\label{eq:trivial_examples}
\end{equation}
where $t \in [0,100]$ with a sampling rate of 20 Hz and $\mathcal{N}$ is Gaussian additive noise.
By applying Eq.~\eqref{eq:dimension_complexity_ratio} to the time series in Eq.~\eqref{eq:trivial_examples}, we can suggest dimensions of 2 for $x_1(t)$, 4 for $x_2(t)$, 6 for $x_3(t)$, and 7 for $x_4(t)$ as shown in Fig.~\ref{fig:example_time_series_permutation_complexity_vs_dimension}. 

In comparison to Takens' embedding, for time series $x_2(t)$ dimension $n=2$ would be sufficient, but if this was used for permutation entropy, no increase in complexity could be detected. Additionally, this result suggests an upper bound on the dimension for permutation entropy as $n \approx 9$ as the ratio in Eq.~\eqref{eq:dimension_complexity_ratio} is approximately 0 for dimensions $n>9$. As a rule of thumb from this result, a dimension of $8$ would be suitable for almost all applications, but it would be optimal to minimize the dimension to reduce the computation time of PE. In Section~\ref{sec:results} we will show the resulting suggested dimensions using this method for a wide variety of dynamical systems.

\section{Results}
\label{sec:results}
This section provides the results of the parameter selection methods. First, in Section~\ref{ssec:results_systems}, we calculate the delay parameter for a wide variety of dynamical systems and data sets using mutual information and the the automatic TDA-based methods described in this manuscript. 
Unfortunately, the optimal parameters cannot be decided based on a simple entropy value comparison since there is no direct equivalence between PE and other entropy approximations of a signal such as Kolmogorov-Sinai (KS) entropy with only a bounding between the two as  $\rm KS \leq PE$~\cite{Keller2017}.
Therefore, to determine the accuracy of the automatically selected PE parameters we implement two other methods of comparison. The first is a comparison to expert suggested parameters for a wide variety of systems (see Section~\ref{ssec:results_systems}). The second approach is a comparison to optimal parameters based on having a significant difference between the PE of two different states for each system. Of course the second method has the requirement that we have a system model or data set with two different states for comparison, which is not typically the case, but does allow for an approximation of optimal PE parameters for these systems. These comparisons are discussed in Section~\ref{ssec:results_systems}. 

The second half of the results, in Sections~\ref{ssec:results_noise}~and~\ref{ssec:results_signal_length}, is based on analyzing the robustness of the automatic TDA-based PE parameter selection methods to additive noise and signal length, respectively.

\subsection{Parameter Value Comparison for Common Dynamical systems} \label{ssec:results_systems}
\begin{figure}[h] 
    \centering
    \includegraphics[width=0.6\textwidth]{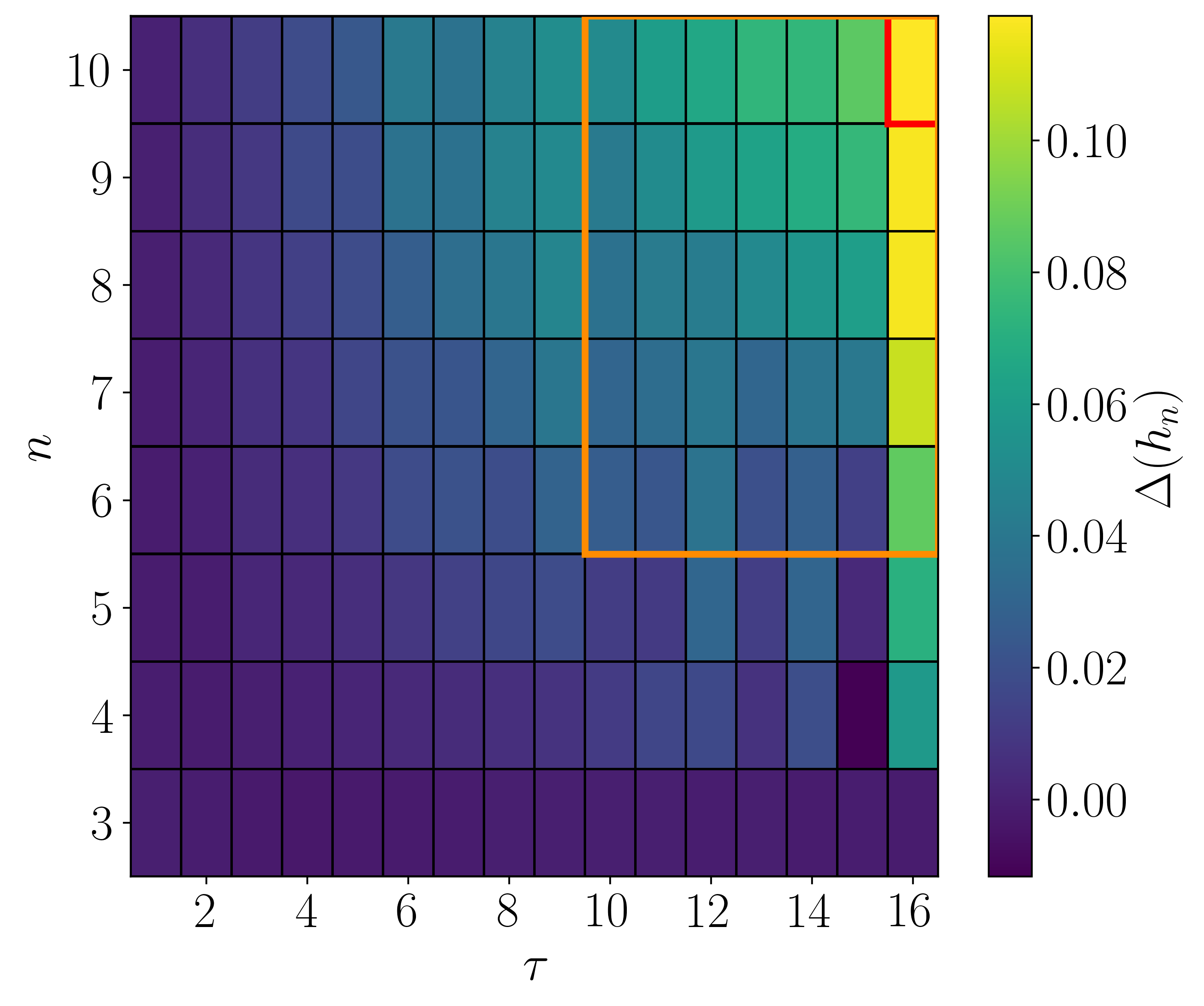}
    \caption{Example showing difference in PE (see Eq.~\eqref{eq:delta_PE}) for periodic and chaotic dynamic states of the Rossler system for a wide range of PE parameters.}
    \label{fig:optimal_params_PE_rossler}
\end{figure}

To determine a range of approximately optimal PE parameters we will quantify the difference between PE values for a wide range of delays and dimensions with the difference for a given $\tau$ and $n$ calculated as
\begin{equation}
\Delta h_n(\tau) = h_n^{({\rm Ch.})}(\tau) - h_n^{({\rm Pe.})}(\tau),
\label{eq:delta_PE}
\end{equation}
where the superscripts Ch. and Pe. represent the PE calculation on the chaotic and periodic time series for the given dynamical system. The specific parameters used to generate periodic and chaotic responses for each system are described in Appendix~\ref{app:systems}.
If we apply Eq.~\eqref{eq:delta_PE} to the Rossler system for $\tau \in [1,15]$ and $n \in [3, 10]$ we find that $\Delta h_n(\tau)$ is significant when $\tau \in [9,15]$ and $n \in [6,10]$ as shown in Fig.~\ref{fig:optimal_params_PE_rossler}. However, as mentioned previously in Section~\ref{sec:dimension}, dimensions greater than 8 can be computationally expensive. 
We consider this range where $\Delta h_n(\tau)$ is relatively large as the range of optimal PE parameters to be compared to. We repeated this process for finding the optimal parameter ranges for PE using a similar procedure to this Rossler example as shown in Table~\ref{tab:PE_parameters_tau}.

\begin{table}[htbp]
\centering
\begin{tabular}{|c|c|c|ccc|c|c|c|c|c|c|c|}
\hline
 &
   &
   &
  \multicolumn{3}{c|}{\textbf{Delay}} &
  \multicolumn{2}{c|}{\textbf{Dim.}} &
  \multicolumn{3}{c|}{} &
  \multicolumn{2}{c|}{} \\
 &
   &
   &
  \multicolumn{2}{c|}{\textbf{\begin{tabular}[c]{@{}c@{}}Sublevel\\ Set Pers.\end{tabular}}} &
   &
  \multicolumn{2}{c|}{\textbf{R}} &
  \multicolumn{3}{c|}{\multirow{-2}{*}{\textbf{\begin{tabular}[c]{@{}c@{}}Expert Suggested\\ Parameters\end{tabular}}}} &
  \multicolumn{2}{c|}{\multirow{-2}{*}{\textbf{\begin{tabular}[c]{@{}c@{}}Opt. Param.\\ Range\end{tabular}}}} \\  
\multirow{-3}{*}{\textbf{Cat.}} &
  \multirow{-3}{*}{\textbf{system}} &
  \multirow{-3}{*}{\textbf{State}} &
  \multicolumn{1}{c|}{\textbf{t}} &
  \multicolumn{1}{c|}{\textbf{f}} &
  \multirow{-2}{*}{\textbf{MI}} &
  \textbf{0.5} &
  \textbf{0.1} &
  \textbf{$\bf \tau$} &
  \textbf{$\bf n$} &
  \textbf{Ref.} &
  \textbf{$\bf \tau$} &
  \textbf{$\bf n$} \\ \hline
 &
  Gauss. &
  - &
  1 &
  1 &
  3 &
  7 &
  8 &
  1 &
  3-6 &
  {\cite{Riedl2013}} &
  - &
  - \\ 
 &
  Uniform &
  - &
  1 &
  1 &
  3 &
  7 &
  8 &
  - &
  - &
  - &
  - &
  - \\ 
 &
  Rayleigh &
  - &
  1 &
  1 &
  2 &
  7 &
  8 &
  - &
  - &
  - &
  - &
  - \\  
\multirow{-4}{*}{\begin{tabular}[c]{@{}c@{}}Noise\\ Models\end{tabular}} &
  Expon. &
  - &
  1 &
  1 &
  2 &
  7 &
  8 &
  - &
  - &
  - &
  - &
  - \\ \hline
 &
   &
  Per. &
  11 &
  7 &
  \cellcolor[HTML]{FFCCC9} 5 &
  5 &
  6 &
   &
   &
   &
   &
   \\ 
 &
  \multirow{-2}{*}{Lorenz} &
  Cha. &
  11 &
  8 &
  11 &
  5 &
  7 &
  \multirow{-2}{*}{10} &
  \multirow{-2}{*}{5-7} &
  \multirow{-2}{*}{\cite{Riedl2013}} &
  \multirow{-2}{*}{8-11} &
  \multirow{-2}{*}{5-10} \\  
 &
   &
  Per. &
   23 &
  15 &
   16 &
  5 &
  6 &
   &
   &
   &
   &
   \\ 
 &
  \multirow{-2}{*}{Rossler} &
  Cha. &
   23 &
  15 &
   18 &
  5 &
  6 &
  \multirow{-2}{*}{9} &
  \multirow{-2}{*}{6} &
  \multirow{-2}{*}{\cite{Tao2018}} &
  \multirow{-2}{*}{15-23} &
  \multirow{-2}{*}{6-10} \\  
 &
   &
  Per. &
  16 &
  \cellcolor[HTML]{FFCCC9}9 &
  14 &
  5 &
  6 &
   &
   &
   &
   &
   \\ 
 &
  \multirow{-2}{*}{\begin{tabular}[c]{@{}c@{}}Bi-direct.\\ Rossler\end{tabular}} &
  Cha. &
  16 &
  13 &
  16 &
  5 &
  6 &
  \multirow{-2}{*}{15} &
  \multirow{-2}{*}{6-7} &
  \multirow{-2}{*}{\cite{Riedl2013}} &
  \multirow{-2}{*}{11-22} &
  \multirow{-2}{*}{6-10} \\ 
 &
   &
  Per. &
  \cellcolor[HTML]{FFCCC9} 2 &
  \cellcolor[HTML]{FFCCC9} 4 &
  \cellcolor[HTML]{FFCCC9} 4 &
  5 &
  6 &
   &
   &
   &
   &
   \\ 
 &
  \multirow{-2}{*}{\begin{tabular}[c]{@{}c@{}}Mackey\\ Glass\end{tabular}} &
  Cha. &
  9 &
  \cellcolor[HTML]{FFCCC9}3 &
  6 &
  5 &
  7 &
  \multirow{-2}{*}{10} &
  \multirow{-2}{*}{4-8} &
  \multirow{-2}{*}{\cite{Zunino2010}} &
  \multirow{-2}{*}{6-12} &
  \multirow{-2}{*}{4-8} \\ 
 &
   &
  Per. &
  17 &
  \cellcolor[HTML]{FFCCC9} 11 &
  \cellcolor[HTML]{FFCCC9} 14 &
  5 &
  6 &
   &
   &
   &
   &
   \\ 
 &
  \multirow{-2}{*}{\begin{tabular}[c]{@{}c@{}}Chua\\ Circuit\end{tabular}} &
  Cha. &
  17 &
  21 &
  17 &
  5 &
  7 &
  \multirow{-2}{*}{20} &
  \multirow{-2}{*}{5} &
  \multirow{-2}{*}{\cite{Shaobo2014}} &
  \multirow{-2}{*}{16-24} &
  \multirow{-2}{*}{5-10} \\  
 &
   &
  Per. &
  4 &
  7 &
  \cellcolor[HTML]{FFCCC9} 2 &
  4 &
  6 &
   &
   &
   &
   &
   \\ 
 &
  \multirow{-2}{*}{\begin{tabular}[c]{@{}c@{}}Coupled \\ Ross.-Lor.\end{tabular}} &
  Cha. &
  4 &
  5 &
  8 &
  5 &
  7 &
  \multirow{-2}{*}{8} &
  \multirow{-2}{*}{3-10} &
  \multirow{-2}{*}{\cite{Staniek2007}} &
  \multirow{-2}{*}{4-8} &
  \multirow{-2}{*}{4-9} \\ 
 &
   &
  Per. &
  \cellcolor[HTML]{FFCCC9} 67 &
  44 &
  7 &
  4 &
  5 &
   &
   &
   &
   &
   \\ 
\multirow{-14}{*}{\begin{tabular}[c]{@{}c@{}}Cont.\\ Flows\end{tabular}} &
  \multirow{-2}{*}{\begin{tabular}[c]{@{}c@{}}Double \\ Pendul.\end{tabular}} &
  Cha. &
  41 &
  28 &
  47 &
  6 &
  7 &
  \multirow{-2}{*}{-} &
  \multirow{-2}{*}{-} &
  \multirow{-2}{*}{-} &
  \multirow{-2}{*}{7-47} &
  \multirow{-2}{*}{5-10} \\ \hline
 &
  Periodic &
  - &
  13 &
  \cellcolor[HTML]{FFCCC9}24 &
  16 &
  4 &
  5 &
  15 &
  4 &
  {\cite{Tao2018}} &
  - &
  - \\  
\multirow{-2}{*}{\begin{tabular}[c]{@{}c@{}}Period.\\ Funct.\end{tabular}} &
  Quasi &
  - &
  25 &
  49 &
  26 &
  6 &
  7 &
  - &
  - &
  - &
  - &
  - \\ \hline
 &
   &
  Per. &
  1 &
  1 &
  3 &
  4 &
  5 &
   &
   &
   &
   &
   \\ 
 &
  \multirow{-2}{*}{Logistic} &
  Cha. &
  1 &
  1 &
  \cellcolor[HTML]{FFCCC9}16 &
  4 &
  6 &
  \multirow{-2}{*}{1-5} &
  \multirow{-2}{*}{4-7} &
  \multirow{-2}{*}{\cite{Riedl2013}} &
  \multirow{-2}{*}{1-4} &
  \multirow{-2}{*}{3-6} \\  
 &
   &
  Per. &
  1 &
  1 &
  3 &
  4 &
  5 &
   &
   &
   &
   &
   \\ 
\multirow{-4}{*}{Maps} &
  \multirow{-2}{*}{Henon} &
  Cha. &
  1 &
  1 &
  \cellcolor[HTML]{FFCCC9}16 &
  6 &
  7 &
  \multirow{-2}{*}{1-2} &
  \multirow{-2}{*}{2-16} &
  \multirow{-2}{*}{\cite{Riedl2013}} &
  \multirow{-2}{*}{1-5} &
  \multirow{-2}{*}{5-8} \\ \hline
 &
   &
  Cont. &
  22 &
  7 &
  17 &
  5 &
  6 &
   &
   &
   &
   &
   \\ 
 &
  \multirow{-2}{*}{ECG} &
  Arrh. &
  15 &
  6 &
  15 &
  5 &
  6 &
  \multirow{-2}{*}{10-32} &
  \multirow{-2}{*}{3-7} &
  \multirow{-2}{*}{\cite{Liu2017}} &
  \multirow{-2}{*}{6-23} &
  \multirow{-2}{*}{5-7} \\ 
 &
   &
  Cont. &
  1 &
  3 &
  6 &
  8 &
  8 &
   &
   &
   &
   &
   \\ 
\multirow{-4}{*}{\begin{tabular}[c]{@{}c@{}}Med.\\ Data\end{tabular}} &
  \multirow{-2}{*}{EEG} &
  Seiz. &
  \cellcolor[HTML]{FFCCC9}12 &
  4 &
  \cellcolor[HTML]{FFCCC9}10 &
  5 &
  7 &
  \multirow{-2}{*}{1-3} &
  \multirow{-2}{*}{3-7} &
  \multirow{-2}{*}{\cite{Riedl2013}} &
  \multirow{-2}{*}{2-6} &
  \multirow{-2}{*}{4-7} \\ \hline
\end{tabular}
\caption{A comparison between the calculated and suggested values for the delay parameter $\tau$. The shaded (red) cells highlight the methods that failed to provide a close match to the suggested delay. }
\label{tab:PE_parameters_tau}
\end{table}

To verify our TDA-based methods for determining $\tau$, Table~\ref{tab:PE_parameters_tau} compares our results to the values from a wide variety of systems for both the first minima of the mutual information function and from expert suggestions, including several listed by Riedl et al.~\cite{Riedl2013}. The table also shows the resulting permutation dimensions suggested from the permutation statistics as described in Section~\ref{sec:dimension} for both $R=0.1$ and $R = 0.5$ from Eq.~\eqref{eq:dimension_complexity_ratio}. For these systems we have also included, where applicable, the delay and dimension parameter estimates for both periodic and chaotic responses to validate each method's robustness to chaos and non-linearity. However, for the medical data section we instead included a healthy/control and unhealthy (arrhythmia for ECG and seizure for EEG) as a substitute for a periodic and chaotic response, respectively. A detailed description of each dynamical system or data set used, including parameters for periodic and chaotic responses, is provided in the Appendix.

In Table~\ref{tab:PE_parameters_tau} we have highlighted the methods that failed to provide an accurate delay $\tau$ in red. This does not mean the embedding is not optimal it simply means it fell outside of the range of expert suggested delays. We will now go through the methods and highlight the advantages and drawbacks as well as provide general suggestions for which method to use based on the category. 

\underline{Noise Models}: We only have one expert suggestion of parameters for the noise models category, which is for Gaussian white noise (Gauss.) as $\tau = 1$ and $n \in [3,6]$. In regards to the delay, all TDA based methods show an accurate selection of $\tau = 1$, however the suggestion of $\tau=3$ from Mutual Information (MI) is slightly higher than suggested. We found that the expert suggested dimensions of 3 to 6 is significantly lower than the minimum dimension suggested by our permutation statistics method of $n=7$. 
As mentioned in Section~\ref{sec:dimension}, we believe it is necessary to have the number of permutations used to be at least less than 50\% of all the permutations available, which corresponds to a dimension $n=7$ for Gaussian noise.
Additionally, as shown in Fig.~\ref{fig:example_time_series_permutation_complexity_vs_dimension}, if a dimension of even $n=6$ is  used approximately 95\% of the permutations are already in use for Gaussian noise, which would make dynamic state changes difficult or statistically insignificant if there is a complexity increase. 
From this logic we can then conclude that a suitable dimension should actually be at least $n=7$ if any increase in the time series complexity is expected. If only decreases in complexity are expected, then a dimension of $n=6$ may be suitable.

\underline{Continuous Flows}: The next category is of continuous flows described by systems of non-linear differential equations. As shown in Table~\ref{tab:PE_parameters_tau}, both the time domain analysis via sublevel set persistence and mutual information provide accurate delay suggestions for many of the examples. We can also conclude that the frequency domain analysis using sublevel set persistence often provided delays that were too small. In regards to the dimension, the suggested dimensions from the permutation statistics agreed with the delay suggested by experts for all of the continuous flow systems. This suggests that our method for selecting a dimension for permutation entropy in Section~\ref{sec:dimension} is accurate for simulations of continuous differential equations.

\underline{Periodic Functions}:
For periodic functions, including a simple sinuisodal function (periodic) and two incommensurate sinuisoidal functions (quasiperiodic), our results in Table~\ref{tab:PE_parameters_tau} show that all methods, including mutual information, provide accurate selections of $\tau$ except the Fourier spectrum analysis via sublevel sets. This method results in a significantly larger suggestion for $\tau$. In regards to the dimension selection, our results using the permutation statistics method described in Section~\ref{sec:dimension} agree with the expert suggested minimum dimension of $n=4$. For the quasiperiodic function tested, there is no reference or expert suggested delay; however, a discrepancy was observed in the obtained time delays where the frequency domain methods yielded a delay of 49 and the time domain and mutual information approaches resulted in delays of about 25. Therefore the results for the quasiperiodic system are inconclusive.

\underline{Maps}:
When selecting the delay parameter for permutations and takens' embedding for maps we found that all of the topological methods suggested accurate delay parameters, while the standard mutual information method selected overly large delay parameters when the maps are chaotic. Therefore, we suggest the use of one of the topological methods when estimating the delay parameter for maps. For the permutation dimension we found a suggested dimension $n \in [4 , 7]$, in comparison to the expected suggested dimension ranging from 2 to 16. While the range suggested from the permutations statistics as described in Section~\ref{sec:dimension} falls within the range suggested by experts, their range is too broad. Specifically, a dimension greater than 9 can be computationally cumbersome, and a dimension lower than 4 would not show significant differences for dynamic state changes. Therefore, we suggest the use of our narrower range of dimension from $n \in [5, 6]$ for maps, which agrees with our optimal PE parameter range.

\underline{Medical Data}:
The medical data used in this study inherently has some degree of additive noise, which provides a first glimpse into the noise robustness of the delay parameter selection methods investigated. However, a more thorough investigation will be provided in Section ~\ref{ssec:results_noise}. From our analysis, we disagree with the expert suggested delay $\tau \in [1,3]$, but rather suggest the delay selected from either mutual information or the time domain analysis of sublevel set persistence. The general selection for delays between 1 and 3 does not account for the large variation in possible sampling rates. If a small delay is used in conjunction with a high sampling rate, an inaccurate delay could be selected resulting in indistinguishable permutation entropy values as the dynamic state changes. In  regards, to the permutation dimension $n$, we believe that a more appropriate dimension, in comparison to the values suggested by experts, should range between 5 and 7 for medical data applications.

\subsection{A Note on Forecasting Performance}
In Table~\ref{tab:PE_parameters_tau}, we show that a significant proportion of the obtained embedding parameters align with expert suggested values. However, some of the systems resulted in parameters outside of these suggested ranges. This section aims to demonstrate that using the embedding parameters from our TDA-based methods can still provide accurate time series forecasting. To perform the forecasting, an AutoRegressive (AR) model was used to learn from simulation data and fit a model of the form,
\begin{equation}
    x_t = \sum_{i=1}^p \varphi_i x_{t-i} + \varepsilon_t,
\end{equation}
where $\varphi_i$ are learned coefficients from the simulation data, $p$ is the order of the model and $\varepsilon_t$ is Gaussian white noise \cite{shumway2000time}. Note also that more sophisticated forecasting models also exist such as the AutoRegressive Moving Average (ARMA) model \cite{shumway2000time}, but for this analysis we consider the simplest such case to demonstrate the accuracy of the estimated embedding parameters and in practice these complex models were found to learn significantly slower compared to the AR model. We assume that we only have access to one state of the system and embed the signal using the appropriate delay and dimension. The embedded signal is then regarded as the ground truth because in practice we do not have access to the remaining states, and Takens' theorem only guarantees a topological equivalence in the reconstructed attractor.

We then fit the autoregressive model to the time series choosing a sufficiently large model order and embed the forecasted signal over a range of delays. These reconstructed attractors are then compared with the embedded attractor from the simulation using the sum of the Root Mean Square Errors (RMSE) across all embedded states of the system and plotted with respect to the embedding delay $\tau$ similar to what is done in \cite{Kantz2003}. If the embedding delay is sufficient for forecasting, we expect that its error will be reasonably close to the error when using the expert suggested delay.

\subsubsection{Periodic Bi-Directional Rossler System}
Using the sublevel persistence method in the frequency domain, the delay for the periodic bi-directional Rossler system was computed to be $\tau=9$ while the expert suggested delay is $\tau=15$. This system was simulated using the specifications in Sec.~\ref{app:bi_rossler} and the AR model was used to forecast the embedded system states between $t=930$ to $t=980$. The AR model was trained on simulation data at an initial condition of $x_1=-0.4$, $y_1=0.6$, $z_1=5.8$, $x_2=0.8$, $y_2=-2.0$, $z_2=-4.0$, and the model was tested by changing $x_1$ to 1.0 and $y_1$ to 0.4. This process was performed for $\tau$ values between 1 and 25 to generate a RMSE error plot with respect to the system delay and a dimension of $n=6$ was used for embedding. The forecast error for this system is shown in Fig.~\ref{fig:forecast_err}(a) where we see that the error for the two delays of interest are relatively close with the sublevel persistence method giving an approximately $10\%$ larger forecast error compared to the expert suggested value. This implies that our method can provide embedding parameters that give accurate forecasting results for periodic signals. We also see that the forecast error for some small delays can be low so it is important to emphasize that this plot does not provide a measure of the optimal delay for attractor reconstruction. 

\subsubsection{Periodic Chua Circuit}
\begin{figure}[htbp]
    \centering
    \begin{minipage}[t]{0.45\textwidth}
        \centering
        \includegraphics[width=\linewidth]{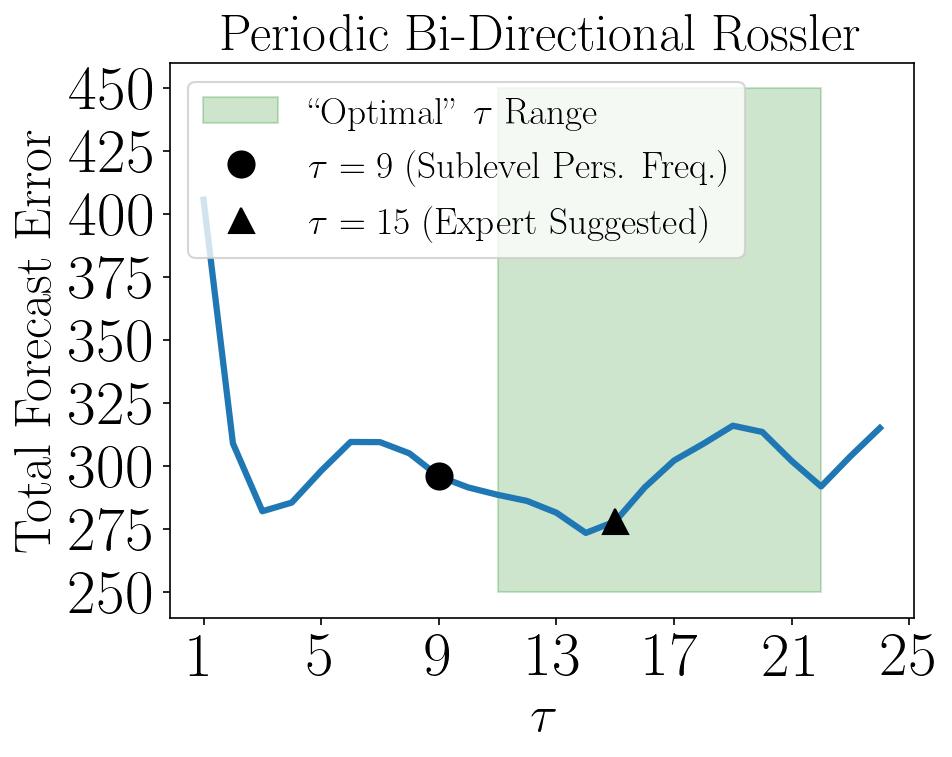} 
        \centering{(a)}
    \end{minipage}
    \hspace{10mm}
    \begin{minipage}[t]{0.45\textwidth}
        \centering
        \includegraphics[width=\linewidth]{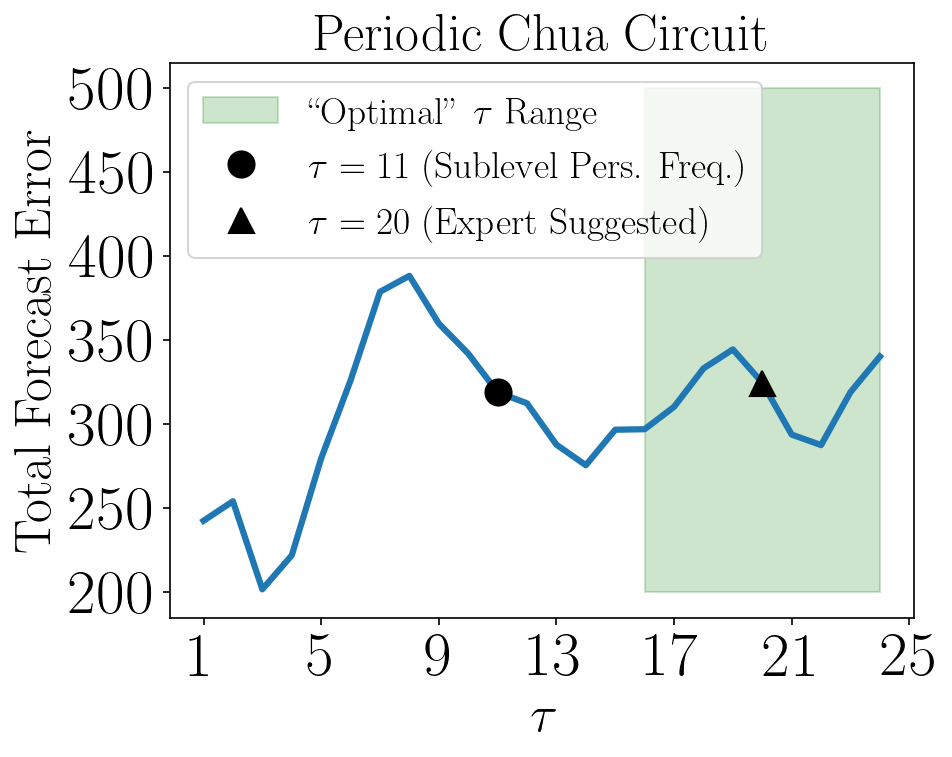} 
        \centering{(b)}
    \end{minipage}
    
    \caption{RMSE forecast error plot for the bi-directional periodic rossler system (a) and periodic chua circuit (b) with respect to the embedding delay using an AR forecast model with 300 terms to estimate the embedded system states. The emphasized values and ranges in the plot are from Table~\ref{tab:PE_parameters_tau}.}
    \label{fig:forecast_err}
\end{figure}

The optimal delay using sublevel persistence with the frequency domain method for the periodic chua circuit was determined to be $\tau=11$ while the expert suggested delay is $\tau=20$. Autoregressive model forecasting was also applied to this system to demonstrate that our methods provide parameters that result in similar forecasting errors compared to expert suggested parameters. We trained a 300 term AR model on simulation data as in Sec.~\ref{app:chua}. The model was used to forecast the system states between $t=138$ to $t=180$ with training initial conditions of $x=1.0$, $y=0.0$, $z=0.0$, and testing initial conditions were set by changing $y=1.0$ and $z=1.0$. A dimension of $n=7$ was used for all simulations of this system and the delay was varied between 1 and 25. The RMSE errors for this system are shown in Fig.~\ref{fig:forecast_err}(b) where we see that for this specific initial condition, the total error for the expert suggested parameter forecast is nearly identical to the result from our method, and because both errors are close this gives further evidence that our method is robust to forecasting.

\subsection{Robustness to Additive Noise} \label{ssec:results_noise}
To determine the noise robustness of the delay parameter selection methods investigated in this work we will use an example time series. Specifically, we will use the $x$ solution to the periodic Rossler system as described in Section \ref{app:rossler}. Note that the simulation parameters for these results differ from those in \cite{Tao2018}. We will use additive Gasussian noise $\mathcal{N}(\mu = 0, \sigma^2)$, where $\sigma$ is determined from the Signal-to-Noise Ratio (SNR). The SNR is a measurement of how much noise there is in the signal with units of decibels (dB) and is calculated as
\begin{equation}
	{\rm SNR_{dB}} = 20 \log_{10} \left( \frac{A_{\rm signal}}{A_{\rm noise}} \right),
\label{eq:SNR}
\end{equation}
where $A_{\rm signal}$ and $A_{\rm noise}$ are the Root-Mean-Square (RMS) amplitudes of the signal and additive noise, respectively. If we manipulate Eq.~\eqref{eq:SNR} we can solve for $A_{\rm noise}$ as
\begin{equation}
	A_{\rm noise} = A_{\rm signal} {10}^{-\frac{{\rm SNR_{dB}}}{20}}.
\label{eq:A_noise}
\end{equation}
Because $x(t)$ is a discrete sampling from a continous system with $t = [t_1, t_2, \ldots, t_N]$, we calculate $A_{\rm signal}$ as
\begin{equation}
	A_{\rm signal} = \sqrt{\frac{1}{N} \sum_{i=1}^{N} {\left[ x(t_i) -  \bar{x} \right]}^2},
\label{eq:A_signal}
\end{equation}
where $\bar{x}$ is the mean of $x$ and is subtracted from $x(t)$ to center the signal about zero.
with $A_{\rm noise}$ calculated, we set the additive noise standard deviation as  $\sigma = A_{\rm noise}$.

\begin{figure}[h] 
    \centering
    \includegraphics[width=0.6\textwidth]{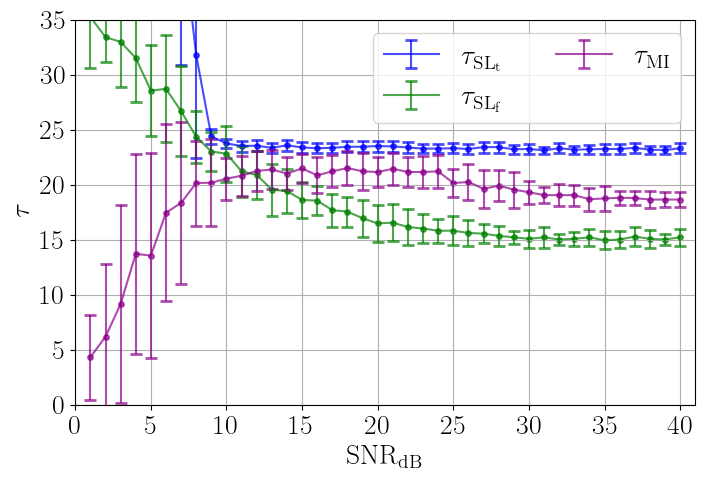}
    \caption{Noise robustness analysis of the delay parameter selection using the Rossler system with incrementing additive noise. The mean and standard deviation as error bars of the delay parameters from 30 trials at each SNR were calculated using sublevel set persistence of the frequency domain $\tau_{\rm SL_f}$, sublevel set persistence of the time domain $\tau_{\rm SL_t}$, and mutual information $\tau_{\rm MI}$.}
    \label{fig:SNR_sweep_tau_methods_rossler_periodic}
\end{figure}

We applied a sweep of the SNR from 1 to 40 in increments of 1 with each SNR being repeated for 30 unique realizations of the noise distributed as $\epsilon \sim \mathcal{N}(0, A_{\rm noise}^2)$. For each realization of $x(t) + \epsilon$ the delay parameters were calculated using all 3 methods: sublevel set persistence of the frequency domain $\tau_{\rm SL_f}$, sublevel set persistence of the time domain $\tau_{\rm SL_t}$, and mutual information $\tau_{\rm MI}$. The mean and standard deviation of the 30 trials at each SNR were calculated for each method as shown in Fig.~\ref{fig:SNR_sweep_tau_methods_rossler_periodic}.
In Fig.~\ref{fig:SNR_sweep_tau_methods_rossler_periodic}, it is clear that the methods are robust to noise down to a SNR of approximately 10 dB. While this does show a limit for the sublevel set persistence methods, SNR values below 10 dB indicate extremely noisy signals.

\subsection{Robustness to Signal Length} \label{ssec:results_signal_length}
\begin{figure}[h] 
    \centering
    \includegraphics[width=0.6\textwidth]{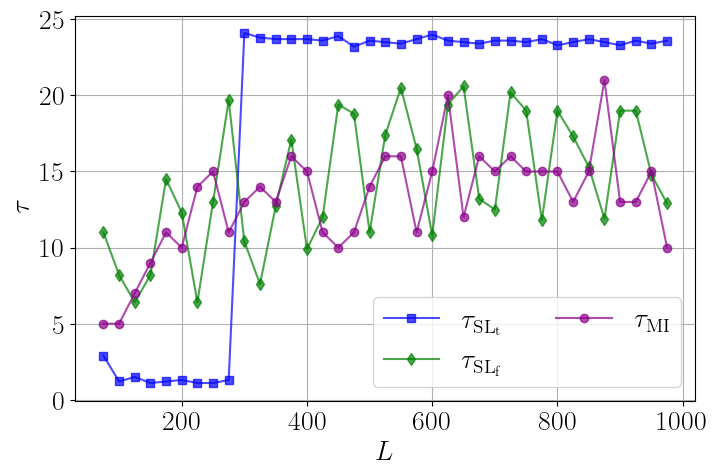}
    \caption{Signal length robustness analysis of the delay parameter selection using the Rossler system with incrementing signal length from 75 to 1000 in steps of 25. The delay parameters were calculated at each $L$ using set persistence of the frequency domain $\tau_{\rm SL_f}$, sublevel set persistence of the time domain $\tau_{\rm SL_t}$ and mutual information $\tau_{\rm MI}$.}
    \label{fig:L_sweep_tau_methods_rossler_periodic}
\end{figure}

A common issue with signal processing and time series analysis methods is their limited functionality with smaller sets of data available, which has been used to analyze the sensitivity of the delay parameter selection~\cite{Deshmukh2020}. Here we will investigate the limitations of these methods in the face of short time series. We will do this analysis by incrementing the length of the time series with the PE parameters calculated at each increment. For our analysis we will again use the Rossler system as described in Section~\ref{app:rossler}. Specifically, we increment the length of the signal from $L=75$ to 1000 in steps of 25 (see Fig.~\ref{fig:L_sweep_tau_methods_rossler_periodic}. However, if this type of analysis is not available for the data set being analyzed, for time series analysis applications it is commonly suggested to have a data length of $L=4000$ for continuous dynamical systems and and $L = 500$ for maps~\cite{Zhang2017}. 

In Fig.~\ref{fig:L_sweep_tau_methods_rossler_periodic} we see that all of the methods reach a larger value of $\tau$, in comparison to the expert suggested $\tau = 9$ when the time series contains at least 300 data points. However, we see that the time domain method seems to be more robust to signal length compared to mutual information and the frequency domain method. An important note to make is that this result is not general for all continuous dynamical systems. The required length of the signal is going to vary significantly depending on the sampling rate of the time series. To determine a general requirement for the methods we repeated this analysis method for all of the systems shown in Table~\ref{tab:PE_parameters_tau}. Our result from this analysis found that, in general, a signal of $L \geq 15 \tau$ allows selecting an appropriate PE and state space reconstruction delay $\tau$ using the TDA-based methods described in this manuscript.

\section{Conclusion}
\label{sec:conclusions}
We described a novel TDA-based approach for automatically determining the PE delay parameter $\tau$ given a sufficiently sampled/over-sampled time series. We investigated the sublevel set persistence in both the time and frequency domains to determine the maximum significant frequency, which was then used to estimate an appropriate delay based on the Shannon-Nyquist sampling criteria~\cite{Melosik2016}.

In regards to the permutation dimension $n$, in Section~\ref{sec:dimension} we developed a simple statistical analysis method for selecting an appropriate permutation dimension $n$ based on the need for only a portion of the permutations to be used in the time series to capture complexity changes. This method also revealed that the permutation dimension and Takens' embedding dimension are not necessarily related and tools for the Takens' embedding dimension cannot generally be used for the permutation dimension. 

To determine the accuracy of these methods, the resulting delays were compared to each of the standard MI approach of Fraser and Swinney \cite{Fraser1986}, expert suggested parameters of various categories of dynamical systems and data sets, and optimal parameters calculated if the dynamical system model allowed for various differing dynamical states to be simulated. This result showed that the sublevel set persistence of the time domain method provided the most accurate delays for all of the systems analyzed except for EEG data. However, due to the differing sampling rates for different EEG data sets, the expert suggested delay of 1--3 may not be accurate for the EEG data investigated in this work. We do not recommend using the sublevel set persistence of the frequency domain as this method consistently provided delays that were too small for continuous flow differential equations and too large for periodic functions. Our noise robustness analysis revealed that the methods we developed here were robust to additive noise up to the SNR values as low as 10 dB. We also analyzed the robustness of the methods to signal length and found they provide an accurate delay even for short time series with the general suggestion of signal length $L > 15 \tau$.

In comparison to the expert suggested dimensions and the optimal dimensions from comparing chaotic and periodic signals, our dimension parameter selection method accurately provided similar results for nearly all of the systems investigated in this work. Additionally, the range suggested using our method was more precise than the dimensions suggested by experts giving the user a more definite answer to an appropriate permutation dimension as these results can vary based on simulation parameters. We hypothesize that using techniques such as a weighted average to combine the parameters from the different methods based on knowledge of the system could lead to even more optimal results. However, this is a topic for future research.

\section*{Acknowledgments}
This material is based upon work supported by the Air Force Office of Scientific Research under award number FA9550-22-1-0007. This work is supported in part by Michigan State University and the National Science Foundation Research Traineeship Program (DGE-2152014) to Max Chumley.

\bibliographystyle{unsrt}  
\bibliography{references}  

\begin{appendices}

\section{Summary of the used data and models} \label{app:systems}
\subsection{Lorenz System}
\label{app:lorenz}
The Lorenz system used is defined as
\begin{equation}
\frac{dx}{dt}   = \sigma (y-x), \: \frac{dy}{dt}   = x (\rho -z) - y, \: \frac{dz}{dt}   = xy - \beta z.
 \label{eq:lorenz}
\end{equation}
The Lorenz system had a sampling rate of 100 Hz. This system was solved for 100 seconds and the last 20 seconds were used. For a periodic response parameters $\sigma = 10.0$, $\beta = 8.0 / 3.0$, and $\rho = 100$ were used. For a chaotic response parameters $\sigma = 10.0$, $\beta = 8.0 / 3.0$, and $\rho = 105$ were used.
\subsection{R\"{o}ssler System}
\label{app:rossler}
The R\"{o}ssler system used was defined as 
   \begin{equation} 
	\frac{dx}{dt}   = -y-z, \: \frac{dy}{dt}   = x + ay, \: \frac{dz}{dt}   = b +z(x-c),
 	\label{eq:rossler}
	\end{equation}
with parameters of $a = 0.10$, $b = 0.20$ for periodic and $b=0.1$ for chaotic, and $c = 14$, which was solved for 1000 seconds with a sampling rate of 15 Hz. Only the last 2500 data points of the solution were used.

\subsection{Coupled R\"{o}ssler-Lorenz System}
\label{app:rossler-lorenz}
The coupled Lorenz-R\"{o}ssler system is defined as
    \begin{equation}
    \begin{split}
	\frac{dx_1}{dt} & = -y_1 - z_1 + k_1 (x_2-x_1), \: \\
	\frac{dy_1}{dt} & =  x_1 + a y_1 + k_2 (y_2-y_1), \:	\\
	\frac{dz_1}{dt} & =  b_2 + z_1 (x_1-c_2) + k_3 (z_2-z_1), \\
	\frac{dx_2}{dt} & = \sigma (y_2-x_2), \\
	\frac{dy_2}{dt} & = \lambda x_2 -y_2 - x_2 z_2, \:	 \\
	\frac{dz_2}{dt} & = x_2 y_2 - b_1 z_2, \\
	\end{split}
 	\label{eq:rossler_lorenz}
	\end{equation}

where $b_1 = 8/3$, $b_2 = 0.2$, $c_2 = 5.7$, $k_1 = 0.1$, $k_2 = 0.1$, $k_3 = 0.1$, $\lambda = 28$, $\sigma = 10$, and $a = 0.25$ for a periodic response and $a = 0.51$ for a chaotic response. This system was simulated at a frequency of 50 Hz for 500 seconds with the last 30 seconds used.

\subsection{Bi-Directional Coupled R\"{o}ssler System}
\label{app:bi_rossler}
The Bi-directional R\"{o}ssler system is defined as
    \begin{equation}
    \begin{split}
	\frac{dx_1}{dt} & = -w_1y_1 - z_1 + k(x_2-x_1), \: \\
	\frac{dy_1}{dt} & = w_1x_1 + 0.165y_1, \:	\\
	\frac{dz_1}{dt} & = 0.2 + z_1(x_1-10), \\
	\frac{dx_2}{dt} & = -w_2y_2 - z2 + k(x_1-x_2), \\
	\frac{dy_2}{dt} & = w_2x_2 + 0.165y_2, \:	 \\
	\frac{dz_2}{dt} & = 0.2 + z_2(x_2-10), \\
	\end{split}
 	\label{eq:rossler_rossler}
	\end{equation}
with $w_1 = 0.99$, $w_2 = 0.95$, $k = 0.25$ for periodic and $k=0.3$ for chaotic. This was solved for 1000 seconds with a sampling rate of 10 Hz. Only the last 150 seconds of the solution were used.

\subsection{Chua Circuit}\label{app:chua}
Chua's circuit is based on a non-linear circuit and is described as
    \begin{equation}
    \begin{split}
	\frac{dx}{dt} & = \alpha (y - f(x)), \: \\
	\frac{dy}{dt} & = \gamma(x - y + z), \:	\\
	\frac{dz}{dt} & = -\beta y, \\
	\end{split}
 	\label{eq:chua}
	\end{equation}
where $f(x)$ is based on a non-linear resistor model defined as
    \begin{equation}
    f(x) = m_1 x + \frac{1}{2}(m_0 + m_1)\left[|x+1| - |x-1|\right].
 	\label{eq:f_resisto}
	\end{equation}

The system parmeters were set to $\beta = 27$, $\gamma = 1$, $m_0 = -3/7$, $m_1 = 3/7$, and $\alpha = 10.8$ for a periodic response and $\alpha = 12.8$ for a chaotic response. The system was simulated for 200 seconds at a rate of 50 Hz and the last 80 seconds were used.

\subsection{Mackey-Glass Delayed Differential Equation}
\label{app:mackey_glass}
The Mackey-Glass Delayed Differential Equation is defined as
    \begin{equation}
	x(t) = -\gamma x(t) + \beta \frac{x(t-\tau)}{1+{x(t-\tau)}^n}
 	\label{eq:Mackey_Glass}
	\end{equation}
with $\tau = 2$, $\beta = 2$, $\gamma = 1$, $n=7.75$ for periodic and $n = 9.65$ for chaotic. This was solved for 400 seconds with a sampling rate of 50 Hz. The solution was then downsampled to 5 Hz and the last 200 seconds were used.

\subsection{Periodic Sinusoidal Function}
\label{app:sine}
The sinusoidal function is defined as
    \begin{equation}
	x(t) = \sin(2\pi t)
 	\label{eq:sinewave}
	\end{equation}
This was solved for 40 seconds with a sampling rate of 50 Hz.

\subsection{Quasiperiodic Function}
\label{app:quasiperiodic}
This function is generated using two incommensurate periodic functions as
\begin{equation}
	x(t) = \sin(\pi t) + \sin(t).
 	\label{eq:quasiperiodic}
\end{equation}
This was sampled such that $t \in [0,100]$ at a rate of 50 Hz.

\subsection{EEG Data}
\label{app:eeg}
The EEG signal was taken from andrzejak et al.~\cite{andrzejak2001}. Specifically, the first 5000 data points from the EEG data of a healthy patient from set A (file Z-093) was used and the first 5000 data points of a patient experiencing a seizure from set E (file S-056) was used.

\subsection{ECG Data}
\label{app:ecg}
The Electrocardoagram (ECG) data was taken from SciPy's misc.electrocardiogram data set. This ECG data was originally provided by the MIT-BIH Arrhythmia Database~\cite{moody2001}. We used data points 3000 to 5500 during normal sinus rhythm and 8500 to 11000 during arrhythmia.

\subsection{Logistic Map}
\label{app:logistic}
The logistic map was generated as
	\begin{equation}
   	x_{n+1} = r x_n(1-x_n),
   	\label{eq:log_map}
	\end{equation}
with $x_0 = 0.5$ and $r = 3.95$. Equation~\ref{eq:log_map} was solved for the first 500 data points.

\subsection{H\'{e}non Map}
\label{app:henon}
The H\'{e}non map was solved as
	\begin{equation}
	\begin{split}
	x_{n+1} & = 1 - a x_n^2 + y_n, \\
	y_{n+1} & = b x_n,
	\end{split}
 	\label{eq:henon_map}
	\end{equation}
where $b = 0.3$, $x_0 = 0.1$, $y_0 = 0.3$, and $a = 1.4$. This system was solved for the first 500 data points of the x-solution.

\subsection{Double Pendulum}
The double pendulum is a staple benchtop experiment for investigated chaos in a mechanical system. A point-mass double pendulum's equations of motion are defined as
	\begin{equation}
	\begin{split}
	\frac{d\theta_1}{dt} & = \omega_1, \\
	\frac{d\theta_2}{dt} & = \omega_2, \\
	\frac{d\omega_1}{dt} & = \frac{N_1}{\ell_1(2m_1 + m_2 - m_2\cos(2\theta_1 - 2\theta_2)}, \\
	\frac{d\omega_2}{dt} & = \frac{N_2}{\ell_2(2m_1 + m_2 - m_2\cos(2\theta_1 - 2\theta_2)}, \\
	N_1 & = 
	-g(2m_1 + m_2)\sin(\theta_1) - m_2h\sin(\theta_1 - 2\theta_2) \\
		& - 2 \sin(\theta_1 - \theta_2)m_2(\omega_2^2 \ell_2 + \omega_1^2 \ell_1\cos(\theta_1 - \theta_2)), \\
	N_2 & = 2\sin(\theta_1-\theta_2)( \omega_1^2 \ell_1 (m_1 + m_2) \\
	&  + g(m_2 + m_2)\cos(\theta_1) + \omega_2^2 \ell_2 m_2 \cos(\theta_1 - \theta_2) ), 
	\end{split}
 	\label{eq:double_pend}
	\end{equation}
where the system parameters $g = 9.81$ $\rm m/s^2$, $m_1 = 1$ kg, $m_2 = 1$ kg, $\ell_1 = 1$ m, and $\ell_2 = 1$ m. The system was solved for 200 seconds at a rate of 100 Hz and only the last 30 seconds were used with initial conditions $[\theta_1, \theta_2, \omega_1, \omega_2] = [0.4{\rm \: rad}, 0.6{\rm \: rad}, 1, 1]$ for periodic and $[\theta_1, \theta_2, \omega_1, \omega_2] = [0, 3{\rm \: rad}, 0, 0]$ for chaotic. This system will have different dynamic states based on the initial conditions, which can vary from periodic, quasiperiodic, and chaotic.

\end{appendices}

\end{document}